\documentclass[superscriptaddress,aps,pra,twocolumn,showpacs,nofootinbib,longbibliography]{revtex4-1}
\usepackage{amsmath}
\usepackage{latexsym}
\usepackage{amssymb}
\usepackage[colorlinks=true,citecolor=blue,urlcolor=blue]{hyperref}
\usepackage{color}
\usepackage{graphics,epstopdf}
\usepackage{soul}
\usepackage[demo]{graphicx}
\usepackage{capt-of}
\usepackage{subcaption}
\usepackage{lipsum}
\usepackage{braket}

\begin{document}

\title{Genuine three qubit Einstein-Podolsky-Rosen steering under decoherence: Revealing hidden genuine steerability via pre-processing}

\author{Shashank Gupta}
\email{shashankg687@bose.res.in}
\affiliation{S. N. Bose National Centre for Basic Sciences, Block JD, Sector III, Salt Lake, Kolkata 700 098, India}
\affiliation{QuNu Labs Pvt Ltd, M. G. Road, Bengaluru, Karnataka 560025, India}

\date{\today}

\begin{abstract}
The behaviour of genuine EPR steering of three qubit states under various environmental noises is investigated. In particular, we consider the two possible steering scenarios in the tripartite setting: (1 $\rightarrow$ 2), where Alice demonstrates genuine steering to Bob-Charlie and (2 $\rightarrow$ 1), where Alice-Bob together demonstrate genuine steering to Charlie.  In both these scenarios, we analyse the genuine steerability of the generalised Greenberger-Horne-Zeilinger (gGHZ) states or the W-class states under the action of noise modelled by amplitude damping (AD), phase flip (PF), bit flip (BF) and phase damping (PD) channels. In each case, we consider three different interactions with the noise depending upon the number of parties undergoing decoherence. We observed that the tendency to demonstrate genuine steering decreases as the number of parties undergoing decoherence increases from one to three. We have observed several instances where the genuine steerability of the state revives after collapsing if one keeps on increasing the damping. However, hidden genuine steerability of a state cannot be revealed solely from the action of noise. So, the parties having a characterised subsystem, perform local pre-processing operations depending upon the steering scenario and the state shared with the dual intent of revealing hidden genuine steerability or enhancing it.
\end{abstract}


\maketitle

\section{Introduction} \label{1}

Quantum networks comprising multiple observers sharing multipartite quantum correlations are the fundamental building blocks of any quantum communication tasks. Depending upon the degree of characterization, a typical multipartite ($>$2) correlations shared among all observers can be of three kinds: genuine entanglement (fully characterized) \cite{KarolZ,Guh}, genuine EPR steering (partially characterized)\cite{He13,Li15,Caval15,Caval16,Riccardi18} or genuine Bell nonlocal (fully uncharacterized) \cite{Coll, Caval11, Ban1}. In real world applications, hybrid quantum networks occurs naturally in the sense that some observers have more knowledge and control of their measuring devices than the others. Genuine EPR steering is present in such hybrid quantum network and is key to realise quantum internet \cite{Kimble2008}.

Understanding genuine EPR steering is foundationally important because detection of such phenomenon certifies the presence of genuine multipartite entanglement in semi-device-independent (SDI) way. Also on the application side genuine multipartite EPR steering is the fundamental resource in several quantum information processing protocols, such as in multipartite secret sharing in a generic SDI scenario \cite{Hil,sss1,sss2}, commercial selftestable quantum key distributions and commercial random number generations \cite{branciard,Mattar2017}. However, similar to the case of other quantum correlations, ubiquitous environmental noise diminishes genuine multipartite EPR steering which inevitably deteriorate the quality of the associated hybrid quantum network.

Extremal or perfect quantum resources (maximize some quantifier) are the most desirable in any quantum information processing task. However, environmental interactions obstruct one from sustaining the perfect quantum correlation, thus, degrading the performance of the implemented task, apart from exceptions in certain special contexts \cite{except1,except2,except3}. One strategy to overcome this practical drawback is local pre-processing by all or some of the parties which reveals the hidden quantum resource \cite{Hirsch13,Tanu19} or enhances it into the perfect quantum resource. In the bipartite context, several pre-processing schemes exist for entanglement \cite{Kim12,Rivu20,Rivu21}, EPR steering \cite{Wang17} and Bell nonlocality \cite{Adel15,Tanumoy19}. A promising pre-processing technique is provided by weak measurements \cite{Vaidman}. The technique of  weak measurement and its reversal has been used to preserve correlations from the effect of amplitude damping channels \cite{Lee11,Man12,Li13,Gupta18}. For multipartite states, pre-processing schemes have also been proposed for entanglement \cite{Ali14,Ali15,Zong15,Kim20} and Bell nonlocality \cite{Chaves12,Chaves14}. In this regard, the genuine EPR steering has recently started gaining attention \cite{Armstrong2015,Wen17,sha20,Liu2020,sha21}.

With this motivation, in this work we analyse the behaviour of genuine steering correlations \cite{He13,Caval15} involving three parties, Alice, Bob and Charlie under the effect of different type of noises. We consider two different genuine steering scenarios where either Alice demonstrates genuine steering to Bob-Charlie (2 $\rightarrow$ 1) or Alice-Bob together demonstrate genuine steering to Charlie (1 $\rightarrow$ 2). The three parties share either a genuinely steerable generalised Greenberger-Horne-Zeilinger (gGHZ) state or a W-class state. Genuine three qubit EPR steering is manifested through the violation of the respective steering inequality in a given scenario \cite{Caval15}. Later, we present the exact pre-processing operations that the parties having a characterised subsystem perform with the dual intent of revealing hidden genuine steerability or enhancing the same. Note that the purpose of pre-processing in this regard is different than the case of entanglement or steering distillation \cite{sha20} where the focus was to concentrate the correlations in multiple imperfect copies to create lesser number of perfect copies whereas in this work, the objective of pre-processing is to reveal the hidden genuine steerability and noise robustness of a single copy of the state.

The paper is organised in the following way. In Sec. \ref{2}, we recapitulate the genuine tripartite steering, the environmental noise modelled by different damping channels and the pre-processing scheme in the steering scenario. In Sec. \ref{3}, we analyse the behaviour of genuine steering in the presence of decoherence. In Sec. \ref{4}, we present the exact pre-processing operations and study the behaviour of genuine steering of the post-processed state under damping. Finally, we conclude this paper with a summary in Sec. \ref{5}.

\section{Overview: Genuine tripartite steering, damping channels and pre-processing operations} \label{2}

Here, we will briefly recapitulate the definition of genuine tripartite steering \cite{He13,Caval15} and various inequalities to detect it. Then, we will overview the action of the damping channels and the pre-processing operations to reveal the hidden genuine steering and to shield the effect of noise.
\subsection{Genuine Tripartite EPR Steering}

Consider that a tripartite state $\rho_{\text{ABC}}$ is shared among three observers: Alice, Bob and Charlie in a quantum network. In the context of tripartite steering, there can be two scenarios: 1) Alice tries to genuinely steer Bob's and Charlie's particles (1$\rightarrow$2 steering) and 2) Alice-Bob try to genuinely steer Charlie's particle (2$\rightarrow$1 steering).

%\subsection{Tripartite Quantum Steering in two-sided device-independent scenario}

In the scenario-1, Alice's measurements are denoted by $X_x$ with outcomes $a$. The POVM elements associated with Alice's measurements are $\{M^A_{a|X_x}\}_{a,x}$ (where $M^A_{a|X_x} \geq 0$ $\forall a, x$; and $\sum_{a} M^A_{a|X_x} = \openone $ $\forall x$). After Alice's measurements, a set of unnormalized conditional states $\{\sigma_{a|X_x}^{\text{BC}}\}_{a,x}$ are formed on Bob-Charli's end given as,

\begin{equation}
	\sigma^{\text{BC}}_{a|X_x} = \text{tr}_A\Big[(\text{M}_{a|X_x}^{A} \otimes \openone_B \otimes \openone_C) \rho_{\text{ABC}} \Big]. % \nonumber \\
						  %&:=& \{ P_{a|x}, \rho_{BC}(a, x) \}_{a \in \{0,...,n_A\}, x \in \{0,...,d_A\}}  
	\label{assemblage1}
\end{equation}
Each element in the assemblage  is given by $\sigma^{BC}_{a|X_x}=p(a|X_x)\varrho_{BC}(a, X_x)$,  where $p(a|X_x)$ is the conditional probability of getting the outcome $a$ when Alice performs the measurement $X_x$; $\varrho_{BC}(a, X_x)$ is the normalized conditional state on Bob's and Charlie's end.

If the state $\rho^{\text{ABC}}$ contains no genuine entanglement then it can be decomposed in the bi-separable form as following 
\begin{eqnarray} \nonumber
	\rho^{\text{ABC}} &=& \sum_\lambda p_\lambda^{A:BC}\rho_\lambda^A \otimes \rho_\lambda^{BC} 
						 + \sum_\mu p_\mu^{B:AC}\rho_\mu^B\otimes \rho_\mu^{AC}\\ \nonumber
						 &+& \sum_\nu p_\nu^{AB:C} \rho_{\nu}^{AB}\otimes \rho_\nu^C,\\ 
\label{bisep}
\end{eqnarray} 
where $p_\lambda^{A:BC}, p_\mu^{B:AC}$ and $p_\nu^{AB:C}$ are probability distributions and $A:BC, B:AC$ and $AB:C$ represent the different types of bipartitions.
In this case, the assemblage (\ref{assemblage1}) has the following form \cite{Caval15}
\begin{eqnarray} 
	\sigma^{\text{BC}}_{a|X_x} &=& \sum_\lambda p_\lambda^{A:BC}p_\lambda(a|X_x)\rho_\lambda^{BC}
							\label{1SA:BC} \\
						 &+& \sum_\mu p_\mu^{B:AC}\rho_\mu^B\otimes \sigma_{a|X_x\mu}^{C} 
						        \label{1SB:AC}. \\ 
						 &+& \sum_\nu p_\nu^{AB:C} \sigma_{a|X_x\nu}^{B}\otimes \rho_\nu^C. \label{1SAB:C} \\
						 \nonumber 
\end{eqnarray} 
Here $p_\lambda(a|X_x)$ denotes the probability of getting the outcome a when Alice performs the measurement denoted by $X_x$ on the state $\rho_\lambda^A$; $\sigma_{a|X_x \mu}^C$ denotes the unnormalized
conditional state on Charlie’s side when Alice gets the
outcome a by performing the measurement denoted by
$X_x$ on the bipartite state $\rho_\mu^{AC}$ shared between Alice and
Charlie; and $\sigma_{a|X_x \nu}^B$ denotes the unnormalized conditional
state on Bob’s side when Alice gets the outcome a by performing the measurement denoted by $X_x$ on the bipartite state $\rho_\nu^{AB}$
shared between Alice and Bob. 

The bi-separable form of the state (\ref{bisep}) imposes constraints on the observed assemblage. For instance, (\ref{1SA:BC}) is an unsteerable assemblage from Alice to Bob-Charlie. The assemblage in (\ref{1SB:AC}) has two features: (i) It is unsteerable from Alice to Bob, but not necessarily from Alice to Charlie; (ii) It is separable. Similarly, the assemblage in (\ref{1SAB:C}) has two features: (i) it is unsteerable from Alice to Charlie, but not necessarily from Alice to Bob; (ii) It is separable. When each element of an assemblage $\{\sigma^{\text{BC}}_{a|X_x}\}_{a,x}$ can be written in the above form, then the assemblage does not demonstrate genuine EPR steering in (1$\rightarrow$2) scenario, otherwise it demonstrates genuine EPR steering in (1$\rightarrow$2) scenario. Now, the question is, given an assemblage ( $\sigma^{\text{BC}}_{a|X_x}$), how to determine whether it is genuine steerable or not? This is a membership problem that can be solved using a semi definite program as the set of genuine steerable assemblages form a convex set and Hahn Banach separation theorem guarantees that a point outside the convex set can be separated from the set using a hyperplane. The solution of the semi definite program in this case are the genuine steering inequalities as mentioned in the coming subsection.%To summarize, any assemblage $\sigma^{\text{BC}}_{a|X_x}$, which does not demonstrate genuine EPR steering, can be expressed as
%\begin{eqnarray} \nonumber 
 %  && \sigma^{\text{BC}}_{a|X_x} = \Gamma_{a|X_x}^{A:BC} + %\Gamma_{a|X_x}^{B:AC} + \Gamma_{a|X_x}^{AB:C}, \\ \nonumber
 %  &&  \Gamma_{a|X_x}^{A:BC} \geq 0,\text{ } \Gamma_{a|X_x}^{B:AC} \geq 0,\text{ } \Gamma_{a|X_x}^{AB:C} \geq 0, \\ \nonumber
 %  && (1) \quad \Gamma_{a|X_x}^{A:BC} \text{ is unsteerable},\\ \nonumber
 %  && (2) \quad \Gamma_{a|X_x}^{B:AC} \text{ is separable and unsteerable from A to B}, \nonumber \\ 
%   && (3) \quad \Gamma_{a|X_x}^{AB:C} \text{ is separable and unsteerable from A to C}.
 %  \label{1SSDP}
%\end{eqnarray}

In scenario-2, Alice's measurements are denoted by $X_x$ with outcomes $a$ and the Bob's measurements are denoted by  $Y_y$  with outcomes $b$. These local measurements by Alice and Bob prepare the assemblage $\{\sigma^{C}_{a,b|X_x,Y_y}\}_{a,b,x,y}$, which are the set of unnormalized conditional states on Charlie's side with 
\begin{equation}
	\sigma^{\text{C}}_{a,b|X_x,Y_y} = \text{tr}_{\text{AB}} \Big[(\text{M}_{a|X_x}^{A} \otimes \text{M}_{b|Y_y}^{B} \otimes \openone_C) \rho_{\text{ABC}} \Big].
	\label{assemblage2}
\end{equation}
Each element in the assemblage  is given by $\sigma^{C}_{a,b|X_x,Y_y}=p(a, b|X_x,Y_y)\varrho_{C}(a, b, X_x, Y_y)$,  where $p(a, b|X_x, Y_y)$ is the conditional probability of getting the outcome $a$ and $b$ when Alice performs the measurement $X_x$ and Bob performs measurement $Y_y$ respectively; $\varrho_{C}(a, b, X_x, Y_y)$ is the normalized conditional state on Charlie's end. 

If the state $\rho^{\text{ABC}}$ contains no genuine entanglement (i.e., it is bi-separable (\ref{bisep})), then the assemblage (\ref{assemblage2}) has the following form \cite{Caval15}
\begin{eqnarray} 
	\sigma^{\text{C}}_{a,b|X_x,Y_y} &=& \sum_\lambda p_\lambda^{A:BC}p_\lambda(a|X_x)\sigma_{b|Y_y\lambda}^{C}
		\label{2SA:BC}\\ 
						 &+& \sum_\mu p_\mu^{B:AC}p_\mu(b|Y_y) \sigma_{a|x\mu}^{C}
						 \label{2SB:AC}\\ 
						 &+& \sum_\nu p_\nu^{AB:C} p_{\nu}(a,b|X_x,Y_y) \rho_\nu^C .
						 \label{2SAB:C} \\ \nonumber
\end{eqnarray} 
Here $p_\lambda(a|X_x)$ denotes the probability of getting the outcome a when Alice performs the measurement denoted by $X_x$ on the state $\rho_\lambda^A$; $\sigma_{b|Y_y \lambda}^C$ denotes the unnormalized
conditional state on Charlie’s side when Bob gets the
outcome b by performing the measurement denoted by $Y_y$ on the bipartite state $\rho_\lambda^{BC}$ shared between Bob and Charlie; $p_\mu(b|Y_y)$ denotes the probability of getting the
outcome b when Bob performs the measurement denoted
by $Y_y$ on the state $\rho_\mu^B$; $\sigma_{a|X_x \mu}^C$ denotes the unnormalized conditional state on Charlie’s side when Alice gets the
outcome a by performing the measurement denoted by
$X_x$ on the bipartite state $\rho_\mu^{AC}$
shared between Alice and Charlie; $p_nu(a,b|X_x,Y_y)$ denotes the joint probability of getting
the outcomes a and b when Alice and Bob perform the
measurements denoted by $X_x$ and $Y_y$, respectively, on the
shared bipartite state $\rho_\nu^{AB}$ between Alice and Bob.

The fact that the state $\rho^{\text{ABC}}$ is bi-separable imposes constraints on the  observed assemblage. For instance, the assemblage (\ref{2SA:BC}) is an unsteerable assemblage from Alice to Charlie, but not necessarily from Bob to Charlie. Similarly, the assemblage (\ref{2SB:AC}) is unsteerable from Bob to Charlie, but not necessarily from Alice to Charlie. The assemblage (\ref{2SAB:C}) has two features: (i) It is unsteerable from Alice-Bob to  Charlie,  (ii) The probability distribution $p_{\nu}(a,b|X_x, Y_y)$ arises due to local measurements performed on a possibly entangled state, it may contain nonlocal quantum correlations. When each element of an assemblage $\{\sigma^{\text{C}}_{a,b|X_x, Y_y}\}_{a,b,x,y}$ can be written in the above form, then the assemblage does not demonstrate genuine EPR steering in (2$\rightarrow$1) scenario, otherwise it demonstrates genuine EPR steering in (2$\rightarrow$1) scenario. Here also, the steering inequalities address the question that whether an assemblage ($\sigma^{\text{C}}_{a,b|X_x, Y_y}$) is genuine steerable or not? %To summarize, any assemblage $\sigma^{\text{BC}}_{a|X_x}$, which does not demonstrate genuine EPR steering, can be expressed as 
%\begin{eqnarray} \nonumber 
%   && \sigma^{\text{C}}_{a,b|X_x,Y_y} = \Pi_{a,b|X_x,Y_y}^{A:BC} + \Pi_{a,b|X_x,Y_y}^{B:AC} + \Pi_{a,b|X_x,Y_y}^{AB:C}, \\ \nonumber
%   &&  \Pi_{a,b|X_x,Y_y}^{A:BC} \geq 0,\text{ } \Pi_{a,b|X_x,Y_y}^{B:AC} \geq 0,\text{ } \Pi_{a,b|X_x,Y_y}^{AB:C} \geq 0, \\ \nonumber
%   && \Pi_{a,b|X_x,Y_y}^{A:BC} \text{ is unsteerable from A to C},\\ \nonumber
 %  && \Pi_{a,b|X_x,Y_y}^{B:AC} \text{ is unsteerable from B to C},\\ \nonumber
%   && \Pi_{a,b|X_x,Y_y}^{AB:C} \text{ is unsteerable from AB to C and}\\ \nonumber
%   && \text{each of the hidden variable probability distributions} \\ \nonumber
 %  && \text{at Alice-Bob's end is quantum correlation produced due} \\
%   && \text{to local measurements}.
%   \label{2SSDP}
%\end{eqnarray}

%Now, If the assemblage (\ref{assemblage2}) can be decomposed in the bi-separable form which means it satisfies the conditions (\ref{2SSDP}) then it is not a genuinely steerable assemblage otherwise it is a genuinely steerable assemblage in 2SDI scenario. 

\subsubsection{Genuine tripartite EPR steering inequalities}

Cavalcanti \textit{et al.} designed several inequalities \cite{Caval15} to address the membership problem of the tripartite assemblage using the optimized numerical solution of the well framed semi definite program (SDP) that detect genuine entanglement of GHZ state given by, $\frac{1}{\sqrt{2}}(|000\rangle + |111 \rangle)$ and W state given by, $\frac{1}{\sqrt{3}}(|001\rangle + |010 \rangle + |100\rangle)$ \cite{Acin,wstate} in the two scenarios mentioned above. These inequalities are nothing but genuine EPR steering inequalities \cite{Caval15}. For GHZ state in $1\rightarrow2$ scenario, the inequality has the following form: 
\begin{align}
G_1 &= 1 + 0.1547 \langle Z_BZ_C\rangle - \frac{1}{3} ( \langle A_3Z_B \rangle + \langle A_3Z_C \rangle  \nonumber \\
& + \langle A_1X_BX_C\rangle - \langle A_1Y_BY_C\rangle - \langle A_2X_BY_C\rangle \nonumber \\
& - \langle A_2Y_BX_C\rangle )\geq 0,
\label{GHZ1}
\end{align}
with $A_i$ for $i=1, 2, 3$, being the observables associated with Alice's uncharacterized measurements with outcomes $\pm 1$ and $X$, $Y$ and $Z$ represent Pauli operators. The coefficients like 0.1547 are the result of the optimised numerical solution of the SDP framed to address the tripartite assemblage membership problem. The GHZ state violates the inequality by $-0.845$ when Alice's measurements are $X$, $Y$ and $Z$, which numerical optimisation suggests are the optimal choices for Alice. %For the GGHZ state ($\cos (\theta) \ket{00} + \sin (\theta) \ket{11}$), the left hand side of the inequality is $0.488033\, -2.66667 \sin (\theta ) \cos (\theta )$, where the measurements of all three parties are same as above. This expression is a monotonic function of $\theta$ and has minimum value for $\theta = \frac{\pi}{4}$ (GHZ state). Hence, the inequality $(\ref{GHZ1})$ is a genuine steering quantifier for the GGHZ state in 1SDI scenario. Note that the inequality $(\ref{GHZ1})$ is violated by GGHZ state for $\theta \in \{ 0.185, 1.378\}$ and so it is genuinely steerable in this range of $\theta$ only.

For GHZ state in 2$\rightarrow$1 scenario, the inequality has the following form: 
\begin{align}
G_2 &= 1 - \alpha (\langle A_3B_3\rangle +  \langle A_3Z \rangle + \langle B_3Z \rangle) - \beta( \langle A_1B_1X \rangle \nonumber \\
& -  \langle A_1B_2Y\rangle -  \langle A_2B_1Y\rangle -   \langle A_2B_2X\rangle )\geq 0,
\label{GHZ2}
\end{align}
where $\alpha = 0.183$, $\beta = 0.258$ are the results of the SDP, $A_i$ and $B_i$ with $i=1, 2, 3$ represent the observables associated with Alice and Bob's uncharacterized measurements, respectively, with outcomes $\pm 1$. The optimal quantum violation of this inequality is $-0.582$. This is achieved by GHZ state when Alice and Bob both perform $X$, $Y$ and $Z$ measurements. %For the GGHZ state, the left hand side of the inequality is $0.4507\, -2.0656 \sin (\theta ) \cos (\theta )$, where the measurements of all three parties are same as above. Note that the inequality $(\ref{GHZ2})$ is violated by GGHZ state for $\theta \in \{ 0.22, 1.34\}$ and so it is genuinely steerable in this range of $\theta$ only.

Similar inequalities for W-state are given for both the scenarios. For W state in 1$\rightarrow$2 scenario, the inequality has the following form: 
\small{
\begin{align}
W_1 &= 1 + 0.4405(\langle Z_B \rangle + \langle Z_C \rangle) - 0.0037  \langle Z_BZ_C\rangle   \nonumber \\
& - 0.1570 ( \langle X_B X_C \rangle + \langle Y_B Y_C \rangle + \langle A_3 X_B X_C \rangle + \langle A_3 Y_B Y_C \rangle )  \nonumber \\
& + 0.2424 ( \langle A_3 \rangle +   \langle A_3Z_BZ_C \rangle ) + 0.1848 ( \langle A_3Z_B\rangle + \langle A_3Z_C \rangle )  \nonumber \\
& - 0.2533( \langle A_1X_B\rangle + \langle A_1X_C\rangle + \langle A_2Y_B\rangle + \langle A_2Y_C\rangle  \nonumber \\ 
& + \langle A_1X_BZ_C\rangle + \langle A_1Z_BX_C\rangle ) + \langle A_2Y_BZ_C\rangle + \langle A_2Z_BY_C\rangle ) \nonumber \\
& \geq 0,
\label{W1}
\end{align}}
 the coefficients are the result of the optimised numerical solution of the SDP framed to address the tripartite assemblage membership problem. The  W state achieving the optimal quantum violation $-0.759$. %For the general W state ($c0 \ket{100} + c1 \ket{010} + c2 \ket{001}$, where $c2 = \sqrt{1-c0^2-c1^2}$), the left hand side of the inequality is $0.0229 \text{c0}^2-2.0264 \text{c0} \text{c1}-2.0264 \text{c0} \text{c2}+0.0037 \text{c1}^2-1.256 \text{c1} \text{c2}+0.0037 \text{c2}^2+1.$, where the measurements of all three parties are same as above. This expression is a monotonic function of $c0$ and $c1$ and has minimum value for $c0 = c1 = \frac{1}{\sqrt{3}}$ (W state). Hence, the inequality $(\ref{W1})$ is a genuine steering quantifier for the general W state in 1SDI scenario.

For W state in 2$\rightarrow$1 scenario, the inequality has the following form: 
\begin{align}
W_2 &= 1 + 0.2517(\langle A_3 \rangle + \langle B_3 \rangle) + 0.3520  \langle Z\rangle - 0.1112 ( \langle A_1 X \rangle \nonumber \\
&+ \langle A_2 Y \rangle + \langle B_1 X  \rangle + \langle B_2 Y  \rangle ) + 0.1296 ( \langle A_3 Z \rangle + \langle B_3 Z \rangle) \nonumber \\
& - 0.1943 ( \langle A_1B_1\rangle + \langle A_2B_2 \rangle ) + 0.2277  \langle A_3B_3\rangle \nonumber \\
& - 0.1590( \langle A_1B_1Z\rangle + \langle A_2B_2Z\rangle ) + 0.2228 \langle A_3B_3Z\rangle  \nonumber \\ 
& -0.2298 (\langle A_1B_3X\rangle + \langle A_2B_3Y\rangle + \langle A_3B_1X\rangle + \langle A_3B_2 Y\rangle ) \nonumber \\
& \geq 0,
\label{W2}
\end{align}
the coefficients are the result of the SDP with the  W state achieving the optimal quantum violation $-0.480$. Although such inequalities are derived for specific state (GHZ or W) but the violation of these inequalities by any state implies genuine steering.
\subsection{Action of damping channels}
 
In this work, four kind of damping channels: Amplitude damping (AD), Phase flip (PF), Bit flip (BF), and Phase damping (PD) are considered \cite{Nielsen}. 
The Kraus operators for these damping channels are given in the table (\ref{KraussT}). These operators satisfy completeness relation $\sum_{i=0}^1 K^{i \dagger} K^i = \openone_2$. The decoherence parameter $p$ ($0\leq p \leq 1$) relates to the decoherence probability. Here, the Kraus operators $K_i$ ($i\in\lbrace0,1\rbrace$) are expressed in the eigenbasis of $\sigma_z$. In the  course of our calculation, we consider the damping channel and decoherence strength to be same for all the decohering party, for the sake of simplicity. We now  consider three cases:
\begin{enumerate}
\item \label{c1} Alice's particle is transmitted through damping channel, Bob and Charlie's particle through an ideal noiseless channel. In this case, the initial state $\rho_{ABC}$ after the action of the damping channel on the Alice's side gets transformed as:
\begin{equation}
 \varrho_{ABC}^{d_A} = \sum_{i = 0}^1\Big(K_{d_A}^i \otimes \openone_2 \otimes \openone_2 \Big) \rho_{ABC} \Big(K_{d_A}^{i^{\dagger}} \otimes \openone_2 \otimes \openone_2  \Big)%}{\text{Tr} \Big[ \sum_{j = 0}^1\Big(K_d^j \otimes \openone_2 \otimes \openone_2 \Big) \rho_{ABC} \Big(K_d^{j^{\dagger}} \otimes \openone_2 \otimes \openone_2  \Big) \Big]} 
 \label{Adamp}
\end{equation}
\item \label{c2} Alice's and Bob's particles are transmitted through unconnected damping channels, but with equal strength and Charlie's particle is transmitted through an ideal noiseless channel. In this case, the initial state $\rho_{ABC}$ after the action of the damping channel on Alice's and Bob's side gets transformed as:
\begin{equation}
 \varrho_{ABC}^{d_A d_B} = \sum_{i,j = 0}^1  \Big(K_{d_A}^i \otimes K_{d_B}^j \otimes \openone_2 \Big) \rho_{ABC} \Big(K_{d_A}^{i^{\dagger}} \otimes K_{d_B}^{j^{\dagger}} \otimes \openone_2  \Big)%}{\text{Tr} \Big[ \sum_{j = 0}^1\Big(K_d^j \otimes \openone_2 \otimes \openone_2 \Big) \rho_{ABC} \Big(K_d^{j^{\dagger}} \otimes \openone_2 \otimes \openone_2  \Big) \Big]} 
 \label{ABdamp}
\end{equation}
\item \label{c3} All three parties' particles are transmitted through unconnected damping channels, but with equal strength. In this case, the initial state $\rho_{ABC}$ after the action of the damping channel on all three parties gets transformed as:
\begin{equation}
 \varrho_{ABC}^{d_A d_B d_C} = \sum_{i,j,k = 0}^1  \Big(K_{d_A}^i \otimes K_{d_B}^j \otimes K_{d_C}^k \Big) \rho_{ABC} \Big(K_{d_A}^{i^{\dagger}} \otimes K_{d_B}^{j^{\dagger}} \otimes K_{d_C}^{k^{\dagger}}  \Big)%}{\text{Tr} \Big[ \sum_{j = 0}^1\Big(K_d^j \otimes \openone_2 \otimes \openone_2 \Big) \rho_{ABC} \Big(K_d^{j^{\dagger}} \otimes \openone_2 \otimes \openone_2  \Big) \Big]} 
 \label{ABCdamp}
\end{equation}
\end{enumerate}
After setting-up the above three situations, depending on the steering scenario either Alice or Alice-Bob demonstrate genuine steering to Bob-Charlie or Charlie respectively.
\begin{table}[] 
\resizebox{0.5\textwidth}{!}{\begin{minipage}{0.7\textwidth}
\caption{Krauss operators for the damping channels} 
\label{KraussT}
\centering
\begin{tabular}{|l|l|l|l|l|l|l|l|l|}
\hline
%\multicolumn{5}{|c|}{$\overline{R}_{g}^E$}                                                                                                                                                                                                                                    \\ \hline
\multicolumn{1}{|c|}{Channels}         & \multicolumn{1}{c|}{Krauss operators} \\ \hline %                              & \multicolumn{1}{c|}{Rank 2}                               & \multicolumn{1}{c|}{Rank 3}                               & \multicolumn{1}{c|}{Rank 4}                               \\ \hline
%\multicolumn{1}{|c|}{$\alpha$}  & \multicolumn{1}{c|}{double}  & \multicolumn{1}{c|}{double} & \multicolumn{1}{c|}{double}  & \multicolumn{1}{c|}{double} \\ \hline
   %                                          &                             &                             &                             &                             \\ \hline
AD                                        & $K_{AD}^0 = \begin{pmatrix}
1 & 0\\
0 & \sqrt{1-p}\\
\end{pmatrix}; ~~~ K_{AD}^1 = \begin{pmatrix}
0 & \sqrt{p}\\
0 & 0\\
\end{pmatrix}  $                                   \\ \hline
PF                                        & $K_{PF}^0 = \begin{pmatrix}
\sqrt{1-p} & 0\\
0 & - \sqrt{1-p}\\
\end{pmatrix}; ~~~ K_{PF}^1 = \begin{pmatrix}
\sqrt{p} & 0\\
0 & \sqrt{p}\\
\end{pmatrix}$                                     \\ \hline
BF                                        & $K_{BF}^0 = \begin{pmatrix}
0 & \sqrt{1-p}\\
\sqrt{1-p} & 0\\
\end{pmatrix}; ~~~ K_{BF}^1 = \begin{pmatrix}
\sqrt{p} & 0\\
0 & \sqrt{p}\\
\end{pmatrix}$                             \\ \hline
PD                                        & $K_{PD}^0 = \begin{pmatrix}
1 & 0\\
0 & \sqrt{1-p}\\
\end{pmatrix}; ~~~ K_{PD}^1 = \begin{pmatrix}
0 & 0\\
0 & \sqrt{p}\\
\end{pmatrix}$                               \\ \hline
%1                                          & 0.843                    & 0.484                   & 0.198                    & 0.081                    \\ \hline
\end{tabular}
\end{minipage}}
\end{table}

\subsection{Pre-processing operations}
In the hybrid quantum tripartite network sharing a three-qubit state, Alice's subsystem is always uncharacterized/untrusted and Charlie's subsystem is always fully characterized/trusted. Depending on whether the scenario is 1$\rightarrow$2 or 2$\rightarrow$1, Bob's subsystem is characterized or uncharacterized respectively. Before going to their respective places through a damping channels, the trusted parties apply the local pre-processing operations to reveal hidden genuine steering or to shield the effect of damping. Let us consider the two asymmetric steering scenario one by one

$\bullet$ {\bf Pre-processing in (1 $\rightarrow$ 2) scenario:} In this case, Alice holds the uncharacterized device and Bob-Charlie holds the characterized devices. Bob and Charlie do the local pre-processing and Alices does nothing before undergoing the action of the damping channel. Each of the two parties with characterized devices (Bob and Charlie) perform a dichotomic qubit POVM  $\textbf{G}^{i}:=\{G_{0}^{i},G_{1}^{i}\}$, satisfying $G_{o^u_i}^{i} \ge 0$ $\forall$ $o^u_i \in \{0,1\}$ and $G_0^{i} + G_1^{i} = \openone$ on the shared state $\rho_{ABC}$ and gets an outcome $o^u_i \in \{ 0, 1\}$. Here $i=B$ for Bob's POVM and $i=C$ for Charlie's POVM.

Next, we introduce the notation $P_{o^u_i}^{i} = \sqrt{G_{o^u_i}^{i}}$ such that $G_{o^u_i}^{i} = P_{o^u_i}^{i^{\dagger}} P_{o^u_i}^{i}$. In the above pre-processing task, when Bob gets the outcome $o^u_B$ and Charlie gets the outcome $o^u_C$ contingent upon performing the above POVMs on the shared state $ \rho_{ABC}$, the state gets updated as
\begin{equation}
 \varrho_{ABC}^{o^u_B, o^u_C} = \frac{\Big(\openone_2 \otimes P_{o^u_B}^{B} \otimes P_{o^u_C}^{C}\Big) \rho_{ABC} \Big(\openone_2 \otimes P_{o^u_B}^{B^{\dagger}} \otimes P_{o^u_C}^{C^{\dagger}} \Big)}{\text{Tr} \Big[ \Big(\openone_2 \otimes P_{o^u_B}^{B} \otimes P_{o^u_C}^{C} \Big) \rho_{ABC} \Big(\openone_2 \otimes P_{o^u_B}^{B^{\dagger}} \otimes P_{o^u_C}^{C^{\dagger}} \Big) \Big]} 
 \label{pBCeqn}
\end{equation}
The updated state $\varrho_{ABC}^{o^u_B, o^u_C}$ is then sent through the damping channels. The action of the damping channel on this state in the three cases is then governed by the Eq. (\ref{Adamp}),(\ref{ABdamp}), and (\ref{ABCdamp})

$\bullet$ {\bf Pre-processing in (2 $\rightarrow$ 1) scenario:} In this case, Alice and Bob hold the uncharacterized devices and only Charlie holds the characterized device. Charlie does the local pre-processing and Alices-Bob do nothing before undergoing the action of the damping channel. Charlie performs a dichotomic qubit POVM  $\textbf{G}^{C}:=\{G_{0}^{C},G_{1}^{C}\}$, satisfying $G_{o^u_C}^{C} \ge 0$ $\forall$ $o^u_C \in \{0,1\}$ and $G_0^{C} + G_1^{C} = \openone$ on the shared state $\rho_{ABC}$ and gets an outcome $o^u_C \in \{ 0, 1\}$. 

In the above pre-processing task, when Charlie gets the outcome $o^u_C$ contingent upon performing the above POVMs on the shared state $ \rho_{ABC}$, the state gets updated as
\begin{equation}
 \varrho_{ABC}^{o^u_C} = \frac{\Big(\openone_2 \otimes \openone_2 \otimes P_{o^u_C}^{C}\Big) \rho_{ABC} \Big(\openone_2 \otimes \openone_2 \otimes P_{o^u_C}^{C^{\dagger}} \Big)}{\text{Tr} \Big[ \Big(\openone_2 \otimes \openone_2 \otimes P_{o^u_C}^{C} \Big) \rho_{ABC} \Big(\openone_2 \otimes \openone_2 \otimes P_{o^u_C}^{C^{\dagger}} \Big) \Big]} 
 \label{pCeqn}
\end{equation}
The updated state $\varrho_{ABC}^{o^u_C}$ is then sent through the damping channels and the action of the damping channels is governed by. the Eq. (\ref{Adamp}),(\ref{ABdamp}) and (\ref{ABCdamp}).
\section{Genuine steering in the presence of decoherence} \label{3}

We are interested in creating genuine steerable network comprising three spatially separated parties. For this, an independent source produces a genuinely entangled three qubit state and send the three particles to three parties constituting the network. During this process, depending on the number of particles undergoing decoherence, three cases may arise that we have discussed before and are governed by Eq. (\ref{Adamp}), (\ref{ABdamp}) and (\ref{ABCdamp}). After setting-up the above three situations, depending on the steering scenario either Alice or Alice-Bob demonstrate genuine steering to Bob-Charlie or Charlie respectively.

In this section, we discuss the three cases-\ref{c1}, -\ref{c2} and -\ref{c3} separately. To evaluate the effect of decoherence on genuine three qubit steering, we start with genuine non-maximally entangled three qubit states of two kinds: (1) generalised GHZ state (2) general W-class state shared by Alice, Bob and Charlie. Note here that, the Cavalcanti et al. have derived the genuine steering inequalities for the GHZ state and W-state, but we have observed that the same inequalities also detect the genuine steering of gGHZ and W-class state for specific range of state parameters that we have computed. The three-qubit generalized GHZ (GGHZ) state,
\begin{equation}
	|\psi_{\text{GGHZ}} \rangle = \cos \theta \ket{000} + \sin \theta \ket{111}, \quad 0 < \theta < \frac{\pi}{4},
	\label{GGHZ}
\end{equation}
where, $\lbrace|0\rangle,|1\rangle\rbrace$ form eigenbasis of $\sigma_z$. The inequality (\ref{GHZ1}) is violated by the GGHZ states (\ref{GGHZ}) for $\theta \in ( 0.185, \frac{\pi}{4})$ whereas the inequality (\ref{GHZ2}) is violated for $\theta \in ( 0.22, \frac{\pi}{4})$ and hence genuine steering is ensured in these ranges only. We will later show that after pre-processing, states outside this $\theta$ range become genuine steerable. We call this feature as revealing \textit{hidden genuine steerability}.  

We also consider two type of pure three qubit W-class states as,
\begin{equation}
	|\psi_\text{GW} \rangle = c_0\ket{001} + c_1\ket{010} + \sqrt{1-c_0^2-c_1^2} \ket{100} ,
\label{wclass}
\end{equation}
with $c_0$, $c_1$ being real and $0 < c_0 \leq \dfrac{1}{\sqrt{3}}$, $0 < c_1 < \dfrac{1}{\sqrt{3}}$. We, therefore, have $\dfrac{1}{\sqrt{3}} < \sqrt{1-c_0^2-c_1^2} < 1$. 
 Let $\mathcal{C}_{0,1}$ = $\{c_0, c_1|0 < c_0 \leq \frac{1}{\sqrt{3}}, 0 < c_1 < \frac{1}{\sqrt{3}} \}$. Note that the inequality (\ref{W1}) is violated by the W-class states (\ref{wclass}) for specific ranges of $c_0$ and $c_1$  and, hence, it ensures 1$\rightarrow$2 genuine steering in that range only. We will denote the set of values of $c_0$ and $c_1$, for which the inequality (\ref{W1}) is violated by the W-class states (\ref{wclass}) with $c_0$, $c_1$ $\in$ $\mathcal{C}_{0,1}$, by the notation $\mathcal{C}^{\text{GW}_1}_{0,1}$. Here, $\mathcal{C}^{\text{GW}_1}_{0,1}$ $\subset$ $\mathcal{C}_{0,1}$. %Henceforth, we will only consider $c_0$, $c_1$ $\in$ $\mathcal{C}^{\text{GW}_1}_{0,1}$. 
 The other type of W-class state which is a one parameter W-class state, 
 \begin{equation}
	\ket{\widetilde{\psi}_{\text{GW}}} = d_0 \ket{001} + \sqrt{\frac{1-d_0^2}{2}} \ket{010} + \sqrt{\frac{1-d_0^2}{2}} \ket{100},
	\label{owclass}
\end{equation}
with $d_0$ being real and $0<d_0<\dfrac{1}{\sqrt{3}}$. Hence, we have $\dfrac{1}{\sqrt{3}} < \sqrt{\dfrac{1-d_0^2}{2}} < \dfrac{1}{\sqrt{2}}$. Note that the inequality (\ref{W2}) is violated by the one parameter W-class states (\ref{owclass}) when $d_0$ $\in$ $\Bigg( \dfrac{3}{25}, \dfrac{1}{\sqrt{3}} \Bigg)$ and, hence, it ensures 2$\rightarrow$1 genuine steering in this range. In this case also we will show that after applying pre-processing, hidden genuine steerability of the states beyond this range of state parameter $d_0$ can be revealed. Let us now discuss the action of various damping channels on the gGHZ state and the behaviour of genuine steering of the decohered state in the three damping cases as mentioned before.

\subsection{gGHZ state under decoherence}

Here we will consider the action of different damping channels, one at a time and study the behaviour of genuine steering in the three cases for that particular damping channel when the shared state is a gGHZ state (\ref{GGHZ}). Let us first consider the case when Alice demonstrates genuine steering to Bob-Charlie.

\subsubsection{Alice to Bob-Charlie}
In this case after the action of the local damping channel, Alice demonstrates genuine steering to Bob-Charlie by the violation of the steering inequality (\ref{GHZ1}). We observe the action of various damping channels as follows:

\begin{figure*}[t!]
\resizebox{14cm}{12cm}{\includegraphics{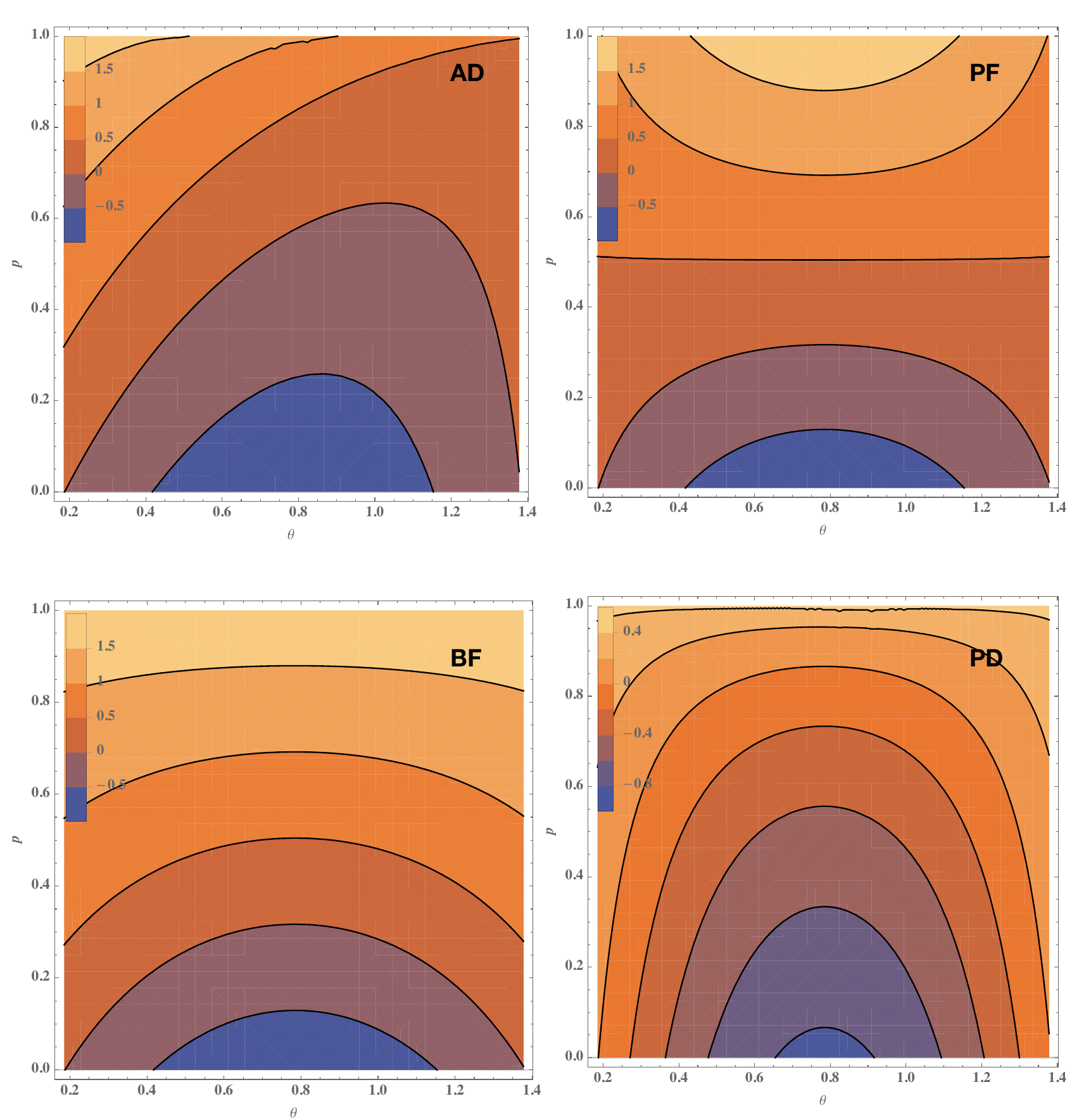}}
\caption{\footnotesize (Coloronline)  Contour plot of genuine steering inequality {\ref{GHZ1}} for gGHZ state in (1 $\rightarrow$ 2) steering scenario versus the damping parameter 'p' (vertical axis) and the state parameter parameter $\theta$ (horizontal axis) in case (\ref{c1}), i.e. when only Alices's particle is undergoing damping. AD, PF, BF and PD stands for amplitude damping, phase flip, bit flip and phase damping channels. The negative value of the inequality implies genuine steering. No signature of revival.}  
\label{fig1}
\end{figure*}
\begin{enumerate}
	\item \textit{Amplitude damping:} We observed that the tendency of the Alice to demonstrate genuine steering to Bob-Charlie decreases as the number of particles undergoing damping increases from one to three as shown in Fig. (\ref{fig1}), (\ref{fig2}) and (\ref{fig3}). Specifically for the $\theta = \pi/3$ gGHZ state, genuine steerability in this case collapses at p = 0.63, 0.47, and 0.35 for the steering scheme (\ref{c1}), (\ref{c2}) and (\ref{c3}) respectively. Interestingly, we have observed that the GHZ state ($\theta = \pi/4$) maximally violates the steering inequality (\ref{GHZ1}) and thus is perfectly genuinely steerable but it is not the most robust state against this noise as the genuine steerability collapses at p = 0.57, 0.45, and 0.34 for the three steering schemes respectively that are lesser than the $\theta = \pi/3$ gGHZ state.
	\item \textit{Phase flip:} As in the case of AD, the tendency of the Alice to demonstrate genuine steering to Bob-Charlie decreases as the number of particles undergoing damping increases from one to three. For the $\theta = \pi/3$ gGHZ state, genuine steerability in this case collapses at p = 0.29, 0.17, and 0.12 for the steering scheme (\ref{c1}), (\ref{c2}) and (\ref{c3}) respectively. Interestingly, we have observed that in case (\ref{c2}), i.e. when only Alice's and Bob's particle undergo damping, the genuine steerability of the state revives after collapsing if one keeps on increasing the damping. For example, for $\theta = \pi/3$ gGHZ state, genuine steerability revives at p = 0.82 after collapsing at p = 0.17. This behaviour is not observed in case (\ref{c1}) and (\ref{c3}). In this case, GHZ state comes out to be most robust state.
	\item \textit{Bit flip:} Behaviour similar to the case of AD and PF is observed when one looks at the tendency of the Alice to demonstrate genuine steering to Bob-Charlie based on the number of particles undergoing damping. For the $\theta = \pi/3$ gGHZ state, genuine steerability in this case collapses at p = 0.27, 0.18, and 0.14 for the steering scheme (\ref{c1}), (\ref{c2}) and (\ref{c3}) respectively. Interestingly, we have observed that in case (\ref{c3}), i.e. when all three particles undergo damping, the genuine steerability of the state revives after collapsing if one keeps on increasing the damping. For example, for $\theta = \pi/3$ gGHZ state, genuine steerability revives at p = 0.86 after collapsing at p = 0.14. This behaviour is not observed in case (\ref{c1}) and (\ref{c2}). In this case also, GHZ state comes out to be most robust state.
	\item \textit{Phase damping:} The tendency of the Alice to demonstrate genuine steering to Bob-Charlie decreases as the number of particles undergoing damping increases from one to three just as in the case of other damping channels. Specifically for the $\theta = \pi/3$ gGHZ state, genuine steerability in this case collapses at p = 0.82, 0.58, and 0.44 for the steering scheme (\ref{c1}), (\ref{c2}) and (\ref{c3}) respectively. In this case also, GHZ state comes out to be most robust state. However, no trace of revival is observed.
\end{enumerate}
In this case, we have observed that the action of phase flip channel in case-(\ref{c2}) an bit flip channel in case-(\ref{c3}) gave rise to revival of the genuine steerable correlations. No signature of revival of the genuine steerable correlation is observed under the action of amplitude damping and phase damping channels. However, we expect that the repeated channel action in asymmetric settings can revive the genuine steerable correlations even under the action of phase damping as well as amplitude damping noise. The analysis is as follows:

{$\bullet$ \textbf{Asymmetric noise action:}} Let the shared state between the three parties is the generalised GHZ state. Depending upon the scenario, either one, two or three parties undergoes first phase damping channel action and then the phase flip channel action with different values of the noise parameters. In the scenario when two parties are interacting in both the channel with the noise parameter in phase damping channel in the range $[0, 0.58)$ and varying strength of the noise in the phase flip channel, we have observed the revival of the genuine steerable correlation. Specifically, the value of the $Eq. (10)$ mentioned in the manuscript is $\frac{1}{3} (-2 \sin ^2(\theta) - 2 \cos^2(\theta)+4 (p_{PD} - 1) (1 - 2 p_{PF})^2 \sin (2 \theta ) + 3.4641)$. Here, $p_{PD}$ and $p_{PF}$ are the noise parameter in the phase damping and phase flip channel respectively. The revival is more visual in the contour plot for the fixed value of $p_{PD} = 0.5$ and varying strength of $p_{PF}$ and $\theta$ as shown in fig. (\ref{pfpd}).

Similar behaviour is observed in the case of amplitude channel followed by phase flip channel as shown in fig. (\ref{pfpd}). We expect similar revival in different scenarios and settings like the action of the bit flip channel after amplitude or phase damping in the case when all three parties undergo damping in both the channels.

\begin{figure}[t!]
\resizebox{8cm}{7cm}{\includegraphics{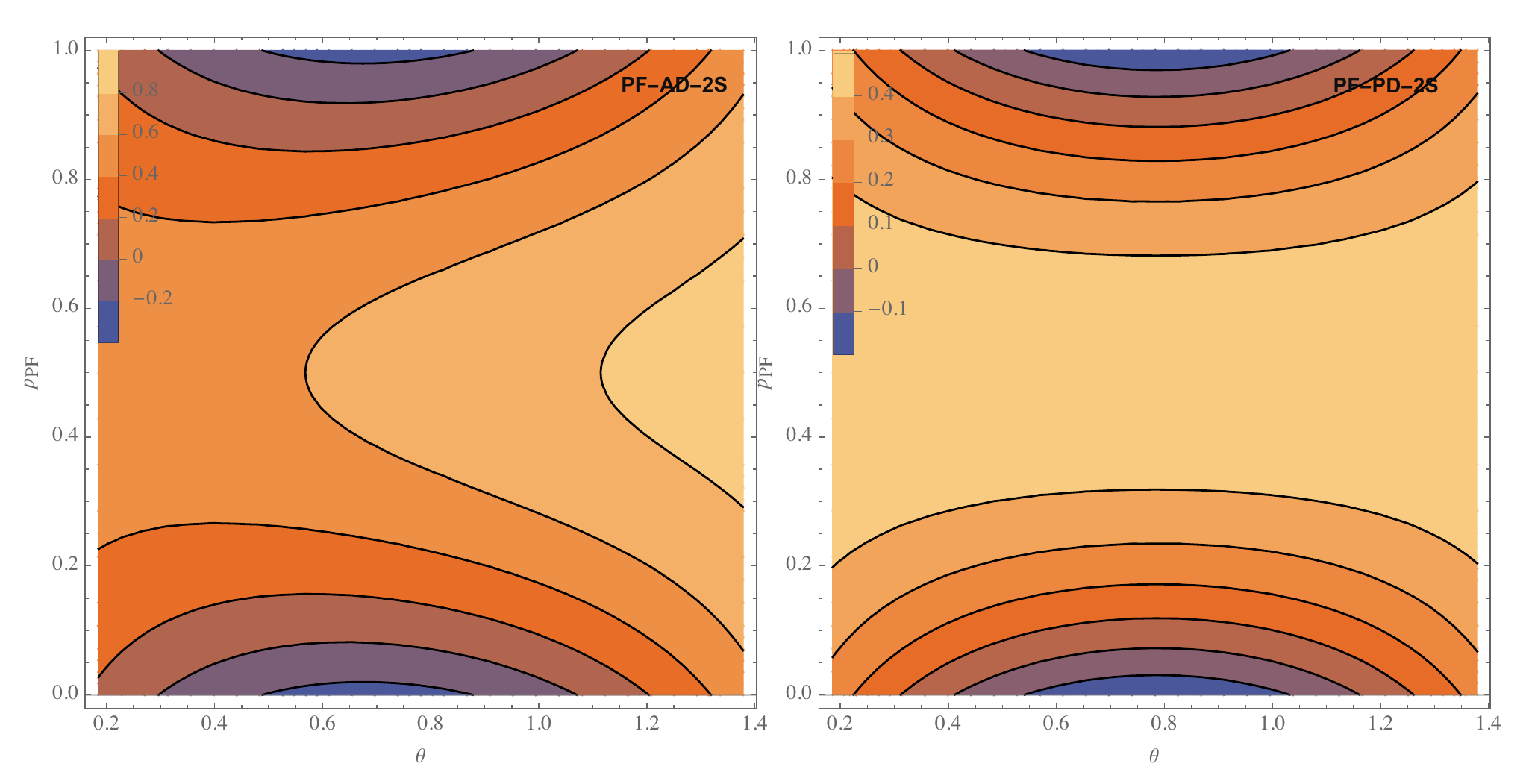}}
\caption{\footnotesize (Coloronline) Contour plot of genuine steering inequality $Eq. (10)$ for gGHZ state in (1 $\rightarrow$ 2) steering scenario versus the damping parameter `$p_{PF}$’ (ordinate) and the state parameter parameter $\theta$ (abscissa) when two parties have interacted first with the phase damping channel of strength $p_{PD} = 0.5$ and then the phase flip channel. The negative value of the inequality implies genuine steering. Genuine steering revives in this case.}  
\label{pfpd}
\end{figure}

Next, consider the case when Alice-Bob demonstrate genuine steering to Charlie.

%\begin{figure}[!ht]
%\resizebox{6cm}{4cm}{\includegraphics{B22.pdf}}
%\caption{\footnotesize (Coloronline) $B^{22}_{max}$ is plotted w.r.t. decoherence parameter $p$ in the scenario of binary input and binary output(2-2). The upper  curve is for case-\ref{c1} when only Alice's part is affected by decoherence. The lower curve is for case-\ref{c2} when both Alice's and Charlie's subsystem interact with the environment. The straight line shows the  upper bound of bilocal correlations.}
%\label{b22}
%\end{figure}
\begin{figure*}[t!]
\resizebox{14cm}{12cm}{\includegraphics{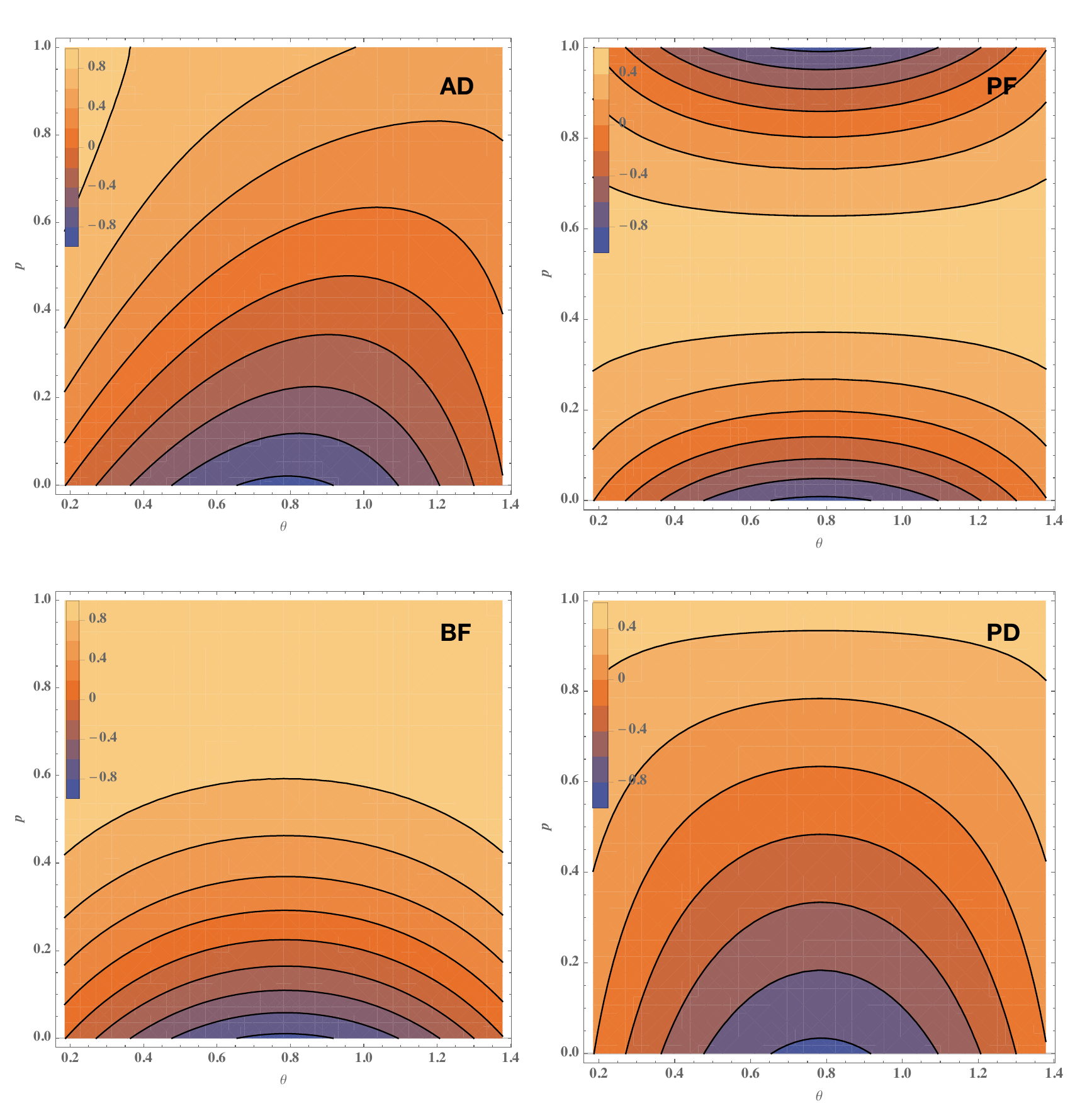}}
\caption{\footnotesize (Coloronline)  Contour plot of genuine steering inequality {\ref{GHZ1}} for gGHZ state in (1 $\rightarrow$ 2) steering scenario versus the damping parameter 'p' (ordinate) and the state parameter parameter $\theta$ (abscissa) in case (\ref{c2}), i.e. when both Alices's and Bob's particles are locally undergoing damping. AD, PF, BF and PD stands for amplitude damping, phase flip, bit flip and phase damping channels. The negative value of the inequality implies genuine steering. Genuine steering revives in case of phase flip damping channel.}  
\label{fig2}
\end{figure*}

\begin{figure*}[t!]
\resizebox{14cm}{12cm}{\includegraphics{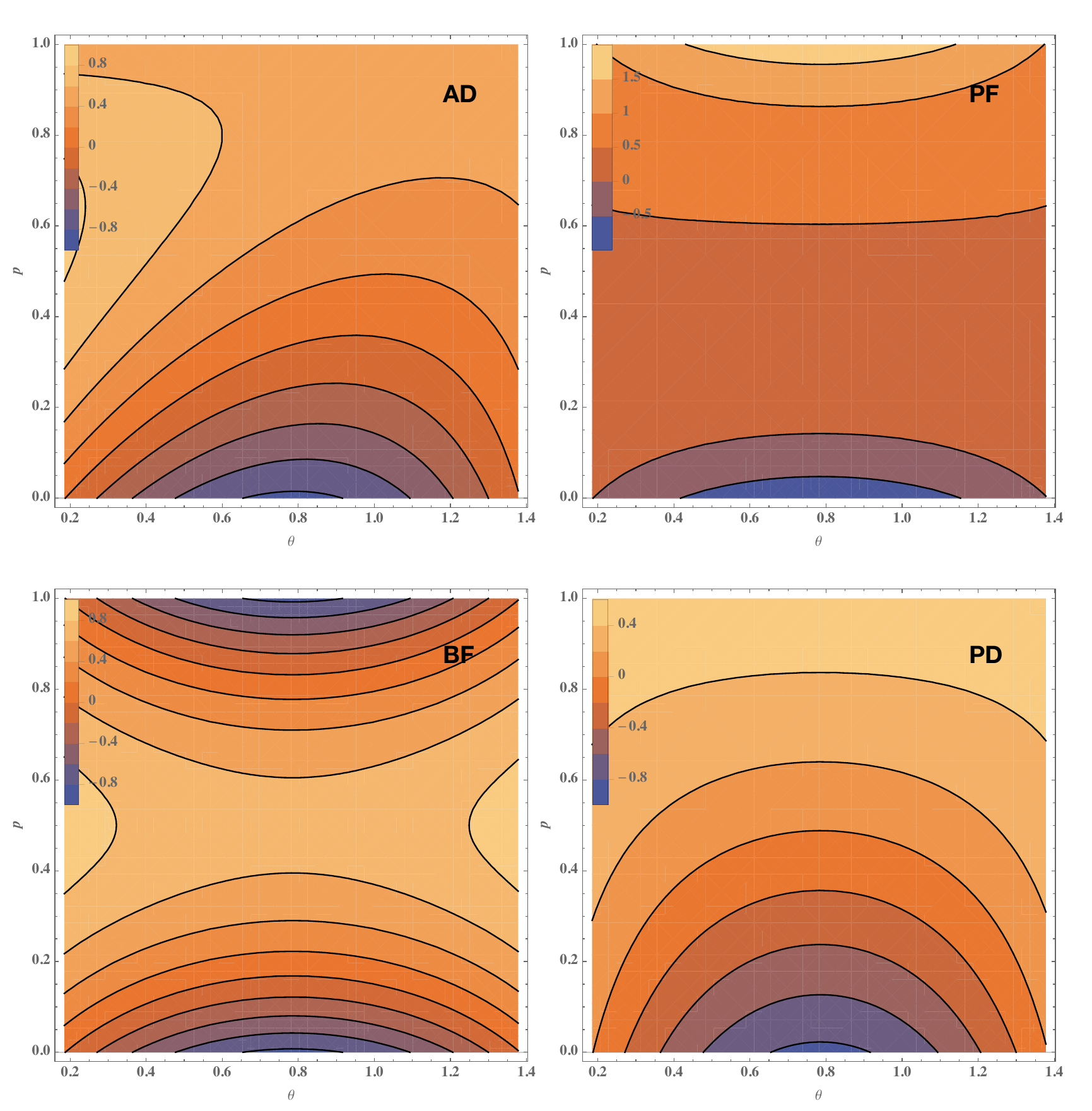}}
\caption{\footnotesize (Coloronline)  Contour plot of genuine steering inequality {\ref{GHZ1}} for gGHZ state in (1 $\rightarrow$ 2) steering scenario versus the damping parameter 'p' (ordinate) and the state parameter parameter $\theta$ (abscissa) in case (\ref{c3}), i.e. when all particles are locally undergoing damping. AD, PF, BF and PD stands for amplitude damping, phase flip, bit flip and phase damping channels. The negative value of the inequality implies genuine steering. Genuine steering revives in case of bit flip damping channel. No signature of revival for other channels.}  
\label{fig3}
\end{figure*}

\subsubsection{Alice-Bob to Charlie}
In this case after the action of the local damping channel, Alice-Bob demonstrate genuine steering to Charlie by the violation of the steering inequality (\ref{GHZ2}). We observe the action of various damping channels as follows:

\begin{figure}[t!]
\resizebox{9cm}{8cm}{\includegraphics{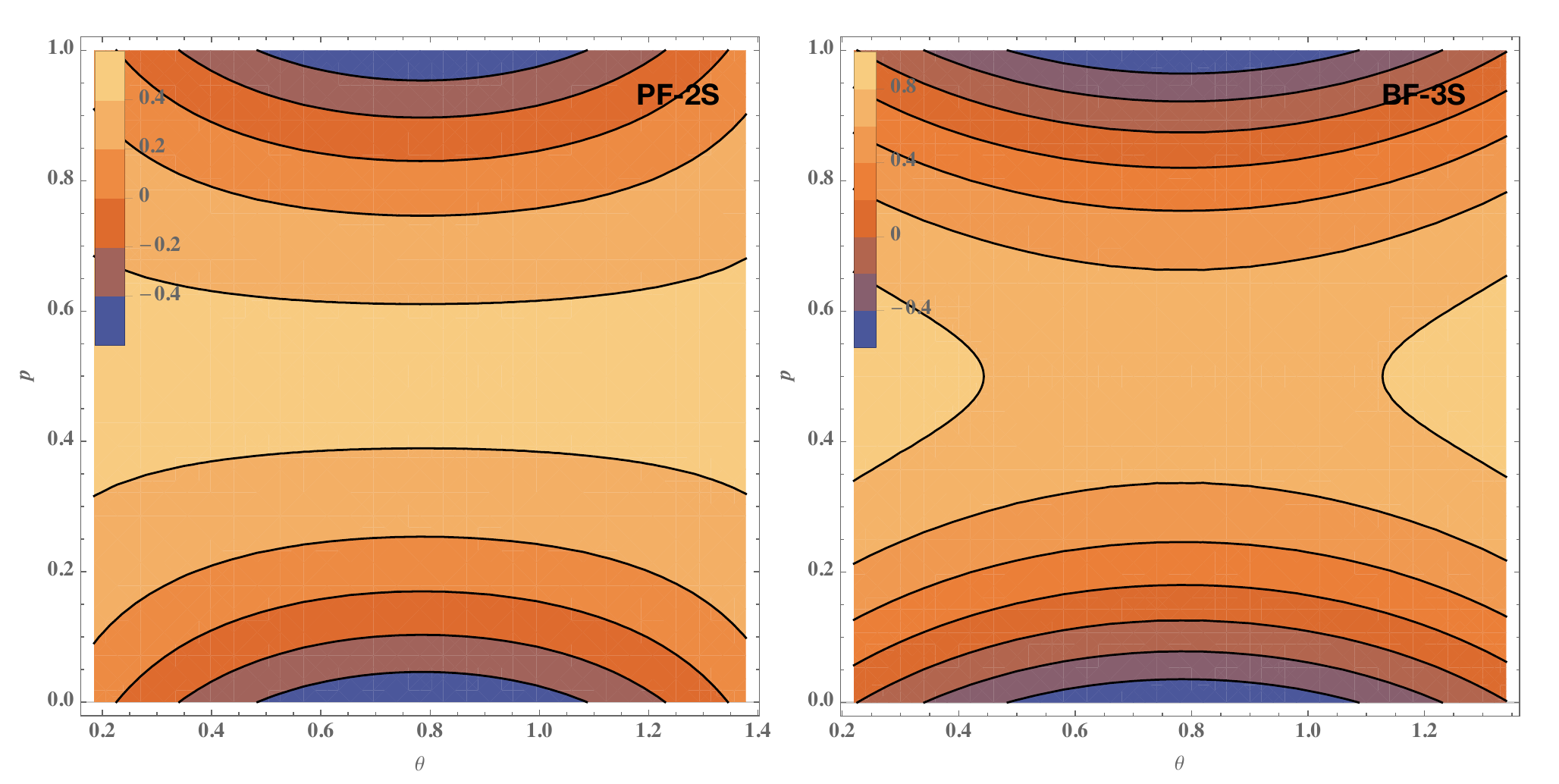}}
\caption{\footnotesize (Coloronline)  Contour plot showing revival of genuine steering inequality {\ref{GHZ2}} for gGHZ state in (2 $\rightarrow$ 1) steering scenario versus the damping parameter 'p' (vertical axis) and the state parameter parameter $\theta$ (horizontal axis). (Left) case (\ref{c2}) and phase-flip noise. (Right) case (\ref{c3}) and bit-flip noise.}  
\label{fig4}
\end{figure}

\begin{enumerate}
	\item \textit{Amplitude damping:} We observed the behaviour similar to the steering scenario when Alice demonstrates genuine steering to Bob-Charlie. Specifically for the $\theta = \pi/3$ gGHZ state, genuine steerability in this case collapses at p = 0.6, 0.37, and 0.27 for the steering scheme (\ref{c1}), (\ref{c2}) and (\ref{c3}) respectively. In this case also GHZ state is not the most robust state although it is the perfectly genuinely steerable.
	\item \textit{Phase flip:} In this case also the tendency of the Alice-Bob to demonstrate genuine steering to Charlie decreases as the number of particles undergoing damping increases from one to three. For the $\theta = \pi/3$ gGHZ state, genuine steerability in this case collapses at p = 0.25, 0.14, and 0.1 for the steering scheme (\ref{c1}), (\ref{c2}) and (\ref{c3}) respectively. Similar to the case Alice demonstrating genuine steering to Bob-Charlie, we have observed that in case (\ref{c2}), the genuine steerability of the state revives after collapsing if one keeps on increasing the damping. For example, for $\theta = \pi/3$ gGHZ state, genuine steerability revives at p = 0.85 after collapsing at p = 0.14 as shown in Fig. (\ref{fig4}). This behaviour is not observed in case (\ref{c1}) and (\ref{c3}). In this case, GHZ state comes out to be most robust state.
	\item \textit{Bit flip:} Behaviour similar to the previous steering scenario is observed. For the $\theta = \pi/3$ gGHZ state, genuine steerability in this case collapses at p = 0.27, 0.15, and 0.1 for the steering scheme (\ref{c1}), (\ref{c2}) and (\ref{c3}) respectively. In this case also, we have observed that in case (\ref{c3}), i.e. when all three particles undergo damping, the genuine steerability of the state revives after collapsing if one keeps on increasing the damping. For example, for $\theta = \pi/3$ gGHZ state, genuine steerability revives at p = 0.9 after collapsing at p = 0.1. This behaviour is not observed in case (\ref{c1}) and (\ref{c2}). In this case also, GHZ state comes out to be most robust state.
	\item \textit{Phase damping:} Behaviour similar to the steering scenario when Alice demonstrate genuine steering to Bob-Charlie is observed. Specifically for the $\theta = \pi/3$ gGHZ state, genuine steerability in this case collapses at p = 0.75, 0.5, and 0.37 for the steering scheme (\ref{c1}), (\ref{c2}) and (\ref{c3}) respectively. In this case also, GHZ state comes out to be most robust state. 
\end{enumerate}
 
We observe that the decohering effect of noise is more pronounced in (2 $\rightarrow$ 1) steering scenario than in (1 $\rightarrow$ 2) scenario.
Next, consider the case when shared state is a W-class state (\ref{wclass}) or one-parameter W-class state (\ref{owclass}).

\subsection{W-class state under decoherence}

Let us analyse the action of different damping channels, one at a time on the genuine steerability in the three cases for that particular damping channel when the shared state is a W-class state (\ref{wclass}). Let us first consider the case when Alice demonstrates genuine steering to Bob-Charlie.

\subsubsection{Alice to Bob-Charlie}
In this case after the action of the local damping channel, Alice demonstrates genuine steering to Bob-Charlie by the violation of the steering inequality (\ref{W1}). We observe the action of various damping channels as follows:
\begin{figure}[t!]
\resizebox{9cm}{8cm}{\includegraphics{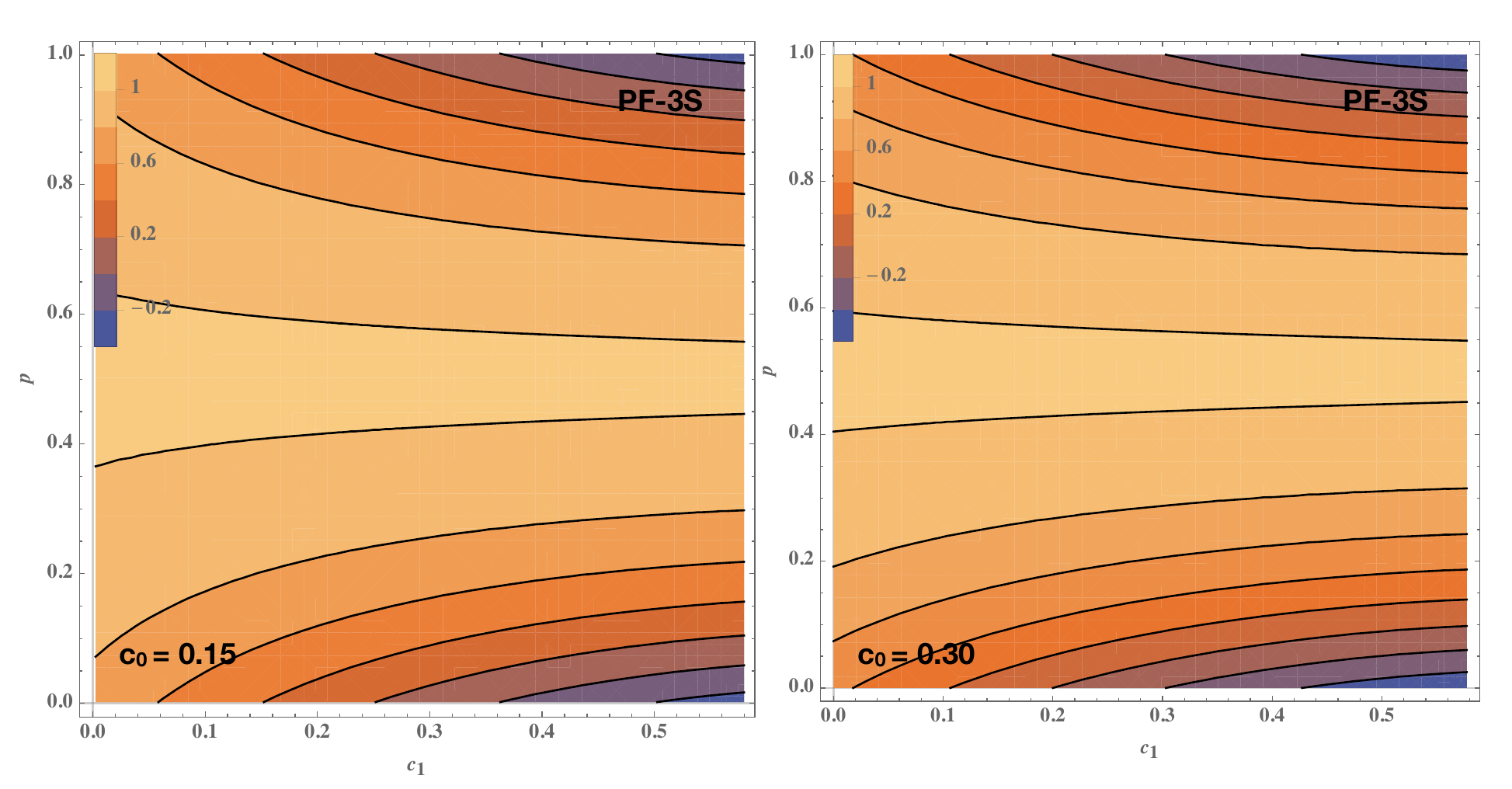}}
\caption{\footnotesize (Coloronline) Contour plot showing revival of genuine steering (using inequality (\ref{W1})) for W-class state (\ref{wclass}) in (1 $\rightarrow$ 2) steering scenario versus the damping parameter 'p' (vertical axis) and the state parameter parameter $c_1$ (horizontal axis) in case-(\ref{c3}) under the action of phase-flip noise. (Left) State parameter $c_0$ is fixed at 0.15. (Right) The state parameter $c_0$ is fixed at 0.3.}  
\label{fig5}
\end{figure}
\begin{enumerate}
	\item \textit{Amplitude damping:} We observed that the general tendency of the Alice to demonstrate genuine steering to Bob-Charlie decreases as the number of particles undergoing damping increases from one to three as in the case when the shared state is a gGHZ state. Specifically for the $c_0 = 0.3$ and $c_1 = 0.5$ W-class state, genuine steerability in this case collapses at p = 0.51, 0.3, and 0.22 for the steering scheme (\ref{c1}), (\ref{c2}) and (\ref{c3}) respectively. Interestingly, we have observed that the W state ($c_0 = 1/\sqrt{3}$ and $c_1 = 1/\sqrt{3}$) maximally violates the steering inequality (\ref{W1}) and thus is perfectly genuinely steerable is also the most robust state against this noise as the genuine steerability collapses at p = 0.61, 0.41, and 0.34 for the three steering schemes respectively that are greater than the $c_0 = 0.3$ and $c_1 = 0.5$ W-class state. Note that it was not the case with the GHZ state.
	\item \textit{Phase flip:} For the $c_0 = 0.3$ and $c_1 = 0.5$ W-class state, genuine steerability in this case collapses at p = 0.18, 0.11, and 0.09 for the steering scheme (\ref{c1}), (\ref{c2}) and (\ref{c3}) respectively. Interestingly, we have observed that in case (\ref{c3}), i.e. when all particles undergo damping, the genuine steerability of the state revives after collapsing if one keeps on increasing the damping. For example, for $c_0 = 0.3$ and $c_1 = 0.5$ W-class state, genuine steerability revives at p = 0.91 after collapsing at p = 0.09. This behaviour is not observed in case (\ref{c1}) and (\ref{c2}). Note that for the gGHZ, revival occurred for the case-(\ref{c2}).
	\item \textit{Bit flip:} For the $c_0 = 0.3$ and $c_1 = 0.5$ W-class state, genuine steerability in this case collapses at p = 0.18, 0.12, and 0.1 for the steering scheme (\ref{c1}), (\ref{c2}) and (\ref{c3}) respectively. Unlike the gGHZ state case, no revival is observed for any steering scheme in this case.
	\item \textit{Phase damping:} For the $c_0 = 0.3$ and $c_1 = 0.5$ W-class state, genuine steerability in this case collapses at p = 0.6, 0.4, and 0.32 for the steering scheme (\ref{c1}), (\ref{c2}) and (\ref{c3}) respectively. 
\end{enumerate}
 Rest of the features are same as in the case of gGHZ state, i.e, W-state comes out to be the most robust state for PF, BF and PD noise and the general tendency of the Alice to demonstrate genuine steering to Bob-Charlie decreases as the number of particles undergoing damping increases from one to three.

Next, consider the case when Alice-Bob demonstrate genuine steering to Charlie.

%\begin{figure}[!ht]
%\resizebox{6cm}{4cm}{\includegraphics{B22.pdf}}
%\caption{\footnotesize (Coloronline) $B^{22}_{max}$ is plotted w.r.t. decoherence parameter $p$ in the scenario of binary input and binary output(2-2). The upper  curve is for case-\ref{c1} when only Alice's part is affected by decoherence. The lower curve is for case-\ref{c2} when both Alice's and Charlie's subsystem interact with the environment. The straight line shows the  upper bound of bilocal correlations.}
%\label{b22}
%\end{figure}

\subsubsection{Alice-Bob to Charlie}
In this case after the action of the local damping channel, Alice-Bob demonstrate genuine steering to Charlie by the violation of the steering inequality (\ref{W2}). We observe the action of various damping channels as follows:

\begin{figure}[t!]
\resizebox{8cm}{7cm}{\includegraphics{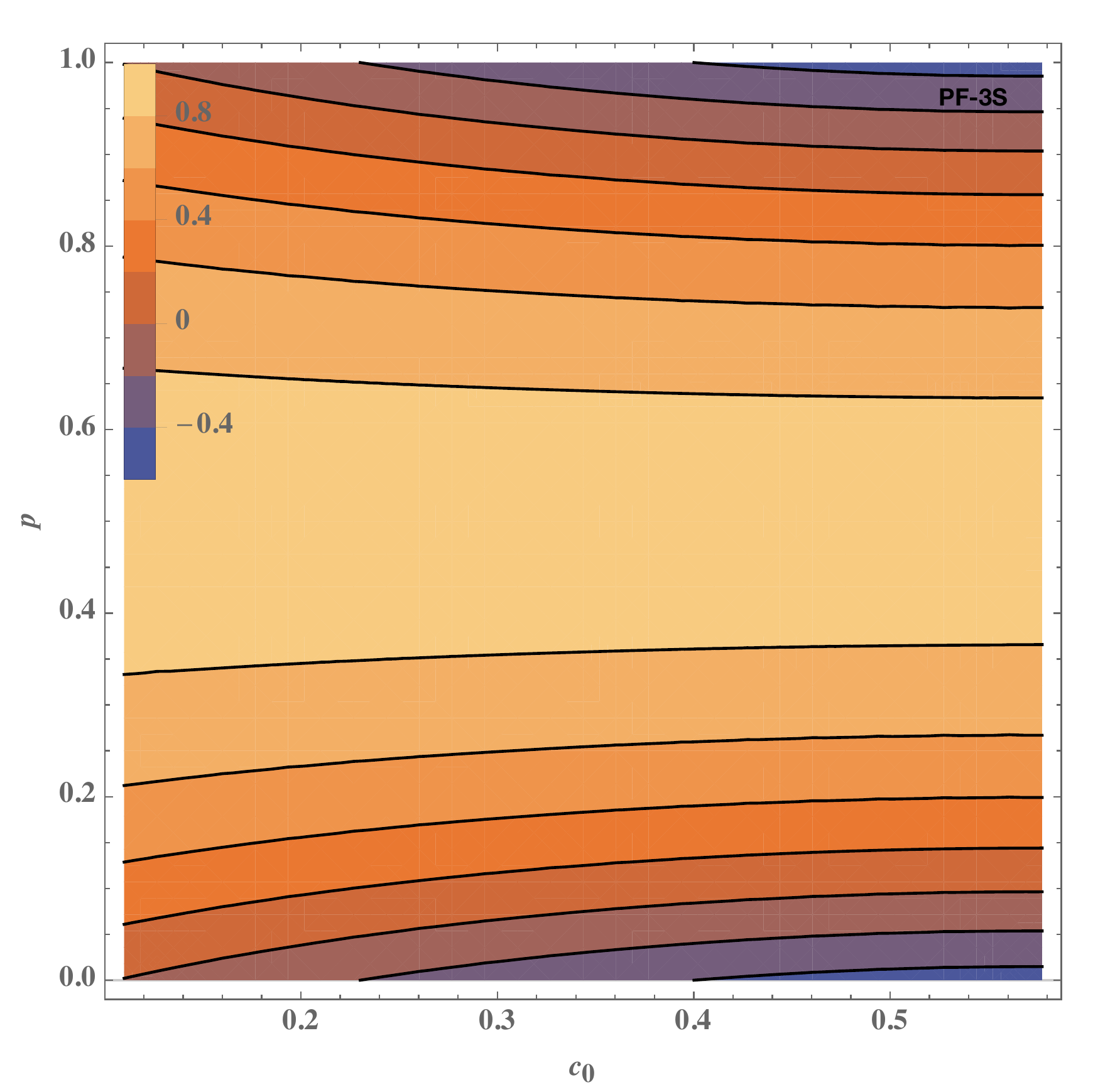}}
\caption{\footnotesize (Coloronline)  Contour plot showing revival of genuine steering (using inequality (\ref{W2})) for W-class state (\ref{owclass}) in (2 $\rightarrow$ 1) steering scenario versus the damping parameter 'p' (vertical axis) and the state parameter parameter $d_0$ (horizontal axis) under the action of phase-flip noise.}  
\label{fig6}
\end{figure}

\begin{enumerate}
	\item \textit{Amplitude damping:} For the $d_0 = 0.3$ W-class state (\ref{owclass}), genuine steerability in this case collapses at p = 0.32, 0.17, and 0.14 for the steering scheme (\ref{c1}), (\ref{c2}) and (\ref{c3}) respectively. 
	\item \textit{Phase flip:} For the $d_0 = 0.3$ W-class state, genuine steerability in this case collapses at p = 0.16, 0.08, and 0.06 for the steering scheme (\ref{c1}), (\ref{c2}) and (\ref{c3}) respectively. Interestingly, in this case also we have observed that when all particles undergo damping, the genuine steerability of the state revives after collapsing. For example, for $d_0 = 0.3$ W-class state, genuine steerability revives at p = 0.93 after collapsing at p = 0.06. This behaviour is not observed in case (\ref{c1}) and (\ref{c2}). Note that for the gGHZ, revival occurred for the case-(\ref{c2}).
	\item \textit{Bit flip:} For the $d_0 = 0.3$ W-class state, genuine steerability in this case collapses at p = 0.14, 0.08, and 0.06 for the steering scheme (\ref{c1}), (\ref{c2}) and (\ref{c3}) respectively.
	\item \textit{Phase damping:} For the $d_0 = 0.3$ W-class state, genuine steerability in this case collapses at p = 0.54, 0.31, and 0.24 for the steering scheme (\ref{c1}), (\ref{c2}) and (\ref{c3}) respectively.
\end{enumerate}
Rest of the features are same as in the steering scenario when Alice demonstrates genuine steering to Bob-Charlie, i.e, W-state comes out to be the most robust state for AD, PF, BF and PD noise and the general tendency of the Alice to demonstrate genuine steering to Bob-Charlie decreases as the number of particles undergoing damping increases from one to three. Also the decohering effect of noise is more pronounced in this scenario than in (1 $\rightarrow$ 2) steering scenario.

Next, we present the pre-processing operations that only the parties with the characterised subsystem performs in order to reveal the hidden genuine steering or to shield the genuine steerability from the different noises for both kind of shared states. Note that motivation of the pre-processing in this case is different that the previous studies like distillation of genuine tripartite steering \cite{sha20} where the focus was to concentrate the genuine steerable correlations contained in multiple imperfect copies of the assemblage to give rise to the lesser number of the perfect assemblages. On the other hand, the motivation for the pre-processing in the present context is 1. Revealing the hidden steerability. 2. Increasing the ronbustness of the state against decoherence at the level of the single copy of the state.

Next, we first consider the specific pre-processing operations for the gGHZ state in the two steering scenario followed by the specific pre-processing operations for two category of the W-states.

\section{Shielding of genuine steering using local pre-processing operations} \label{4}

In the previous section (\ref{3}), we have observed that GHZ state and W-state turn out to be the most robust state against the PF, BF and PD noise. Even in the AD case, GHZ state is more robust than the gGHZ states for $\theta < \pi/4$ as shown in Fig. (\ref{fig1}), (\ref{fig2}) and (\ref{fig3}). So if we perform such pre-processing operations that give a GHZ state from the gGHZ state and W-state from the W-class state probabilistically, then the post-processed state would be more robust than the initial state against the noises under consideration. 

However, in a steering scenario, not all parties subsystem's are characterised. Hence, we are constrained to perform local pre-processing operations only on the parties having characterised subsystems. Depending on the steering scenario (1 $\rightarrow$ 2) or (2 $\rightarrow$ 1), either two or only one party can perform pre-processing operations. Let us now consider the pre-processing operations for gGHZ state and analyse the genuine steerability of the post processed state in two steering scenario under various damping channels.  

\subsection{Pre-processing of gGHZ state}

In this case, the target is to obtain the GHZ state. For this, either Bob and Charlie or only Charlie can perform pre-processing depending on whether Alice or both Alice-Bob wants to demonstrate genuine steering to Bob-Charlie or Charlie respectively. Let us first consider Alice to Bob-Charlie steering scenario.
\subsubsection{Alice to Bob-Charlie}
In this case Bob's and Charlie's subsystems are characterised. There can be two strategies to get the target state depending upon whether one party or both the party pre-process as follows \cite{sha20}.\\
$\textit{Strategy 1: "Equal participation"}$ Both Bob and Charlie participate and perform quantum measurements (POVM) with the following measurement operators on the initial state (\ref{GGHZ}).
\begin{eqnarray}
	P_{0}^{B} &=& \left(
\begin{array}{cc}
\sqrt{ \tan \theta } & 0 \\
 0 & 1 \\
\end{array}
\right), \quad P_{1}^{B} =  \left(
\begin{array}{cc}
 \sqrt{1-\tan \theta } & 0 \\
 0 & 0 \\
\end{array}
\right), \nonumber \\
P_{0}^{C} &=& \left(
\begin{array}{cc}
\sqrt{ \tan \theta } & 0 \\
 0 & 1 \\
\end{array}
\right), \quad P_{1}^{C} =  \left(
\begin{array}{cc}
 \sqrt{1-\tan \theta } & 0 \\
 0 & 0 \\
\end{array}
\right),
\label{GHZsteer1}
\end{eqnarray} 
The target state is a GHZ state when both the parties gets '0' outcome. This can be seen by using Eq. (\ref{pBCeqn}). The probability of success is $\text{Tr}[(\openone_2 \otimes P_0^B \otimes P_0^C)\rho_{gGHZ}(\openone_2 \otimes P_0^{B^{\dagger}} \otimes P_0^{C^{\dagger}})] = 2 \sin^2 \theta$.\\
$\textit{Strategy 2: "Single party participation"}$ Here either Bob or Charlie performs quantum measurement with the following measurement operators on the initial state (\ref{GGHZ})
\begin{equation}
	K_{0}^{B/C} = \left(
\begin{array}{cc}
 \tan \theta  & 0 \\
 0 & 1 \\
\end{array}
\right), \quad K_{1}^{B/C} =  \left(
\begin{array}{cc}
 \sqrt{1 - \tan^2 \theta}  & 0 \\
 0 & 0 \\
\end{array}
\right).
\label{GHZsteer2}
\end{equation}
The target state is a GHZ state when the outcome is '0'. This can be seen by using Eq. (\ref{pCeqn}). The probability of success is $\text{Tr}[(\openone_2 \otimes P_0^B \otimes \openone_2)\rho_{gGHZ}(\openone_2 \otimes P_0^{B^{\dagger}} \otimes \openone_2)] = 2 \sin^2 \theta$ when Bob does pre-processing and it is $\text{Tr}[(\openone_2 \otimes \openone_2 \otimes P_0^C)\rho_{gGHZ}(\openone_2 \otimes \openone_2 \otimes P_0^{C^{\dagger}})] = 2 \sin^2 \theta$ when Charlie does pre-processing. Note that in both the scenario, the success probability is same. Also, with small but finite success probability, we manage to get the GHZ state which are genuine steerable from the gGHZ state ($\theta < 0.185$) that are not genuinely steearble. Hence, using these pre-processing operations we are able to reveal the hidden genuine steerability of the gGHZ states. 

In order to calculate the improvement in the robustness after pre-processing, we will calculate $\delta p_c^g$ which is defined as the difference between the p values of the $\theta = \pi/6$ and $\theta = \pi/4$ gGHZ state at which the genuine steerability of the state collapses. Also, improvement in the revival after pre-processing is observed by $\delta p_r^g$ which is defined as the difference between the p values of the $\theta = \pi/6$ and $\theta = \pi/4$ gGHZ state at which the genuine steerability of the state revives. We now analyse the action of various damping channels on the post-processed state.

\begin{figure}[t!]
\resizebox{8cm}{7cm}{\includegraphics{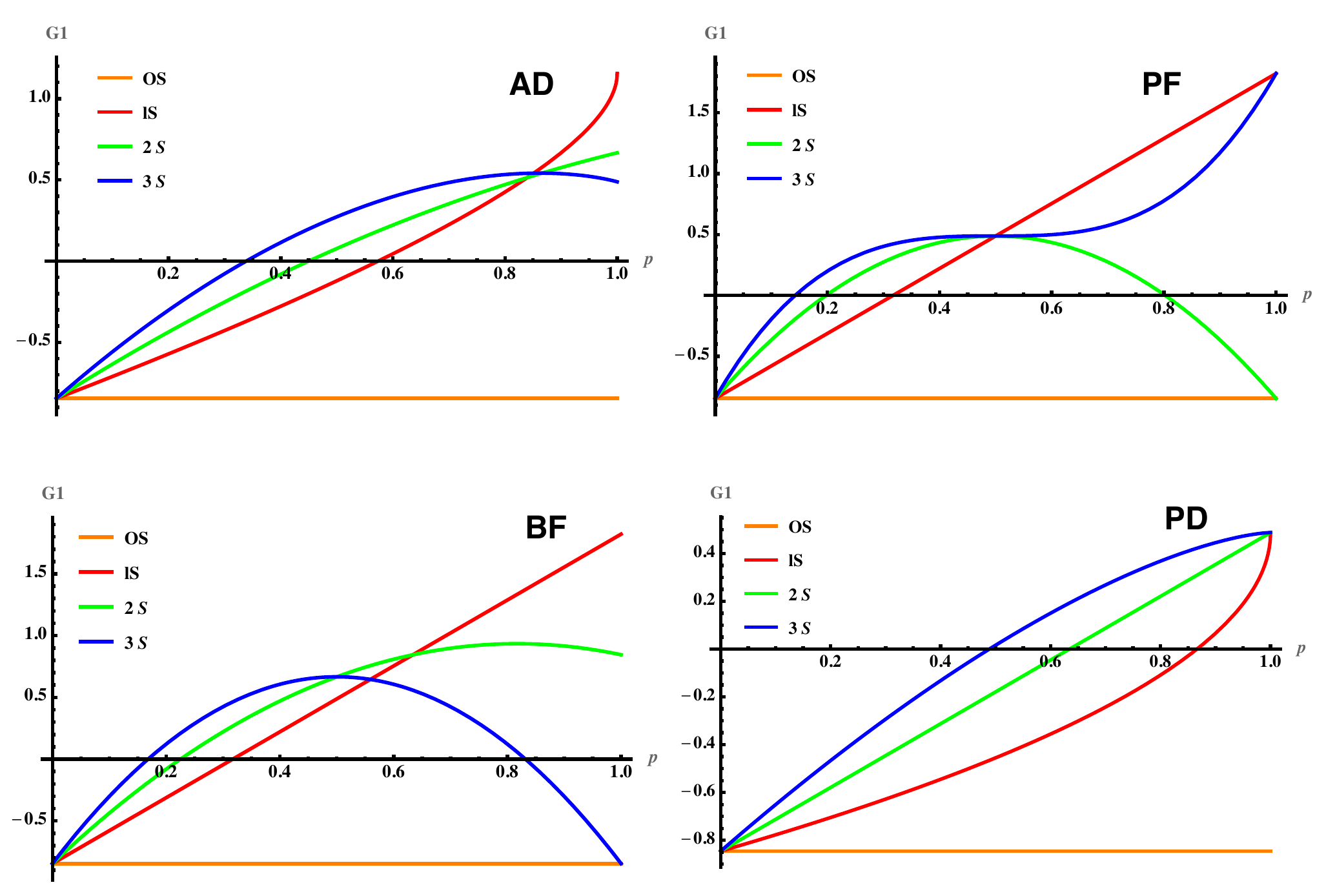}}
\caption{\footnotesize (Coloronline)  The variation of left hand side of the steering inequality (\ref{GHZ1}) (vertical axis) for the post-processed GHZ state in (1 $\rightarrow$ 2) steering scenario versus the damping parameter 'p' (horizontal axis). 0S, 1S, 2S and 3S stands for no damping, one-sided damping as in case-(\ref{c1}), two-sided damping as in case-(\ref{c2}) and all or three-sided damping as in case-(\ref{c3}) respectively.}  
\label{fig7}
\end{figure}

\begin{enumerate}
	\item \textit{Amplitude damping:} We observed similar behaviour as before i.e. the tendency of Alice to demonstrate genuine steering to Bob-Charlie decreases as the number of particles undergoing damping increases from one to three as shown in Fig. (\ref{fig7}).  The values of $\delta p_c^g$ are 0.17, 0.13 and 0.1 for the three cases (\ref{c1}), (\ref{c2}) and (\ref{c3}) respectively. This shows significant improvement in robustness after pre-processing.
	\item \textit{Phase flip:} The values of $\delta p_c^g$ are 0.03, 0.03 and 0.02, the $\delta p_r^g$ is -0.02. The negative sign in $\delta p_r^g$ implies that the genuine steerability of GHZ state revives before than $\theta = \pi/6$ gGHZ state. This implies significant improvement in robustness, as well as, revival after pre-processing.
	\item \textit{Bit flip:} The values of $\delta p_c^g$ are 0.05, 0.03, 0.03 and $\delta p_r^g$ is -0.03 which is a significant improvement in robustness, as well as, revival after pre-processing.
	\item \textit{Phase damping:} In this case, the values of $\delta p_c^g$ are equal to 0.05 for the three cases.
\end{enumerate}

Other features are same as before pre-processing. Next, consider the case when Alice-Bob demonstrate genuine steering to Charlie.

%\begin{figure}[!ht]
%\resizebox{6cm}{4cm}{\includegraphics{B22.pdf}}
%\caption{\footnotesize (Coloronline) $B^{22}_{max}$ is plotted w.r.t. decoherence parameter $p$ in the scenario of binary input and binary output(2-2). The upper  curve is for case-\ref{c1} when only Alice's part is affected by decoherence. The lower curve is for case-\ref{c2} when both Alice's and Charlie's subsystem interact with the environment. The straight line shows the  upper bound of bilocal correlations.}
%\label{b22}
%\end{figure}

\subsubsection{Alice-Bob to Charlie}
\begin{figure}[t!]
\resizebox{8cm}{7cm}{\includegraphics{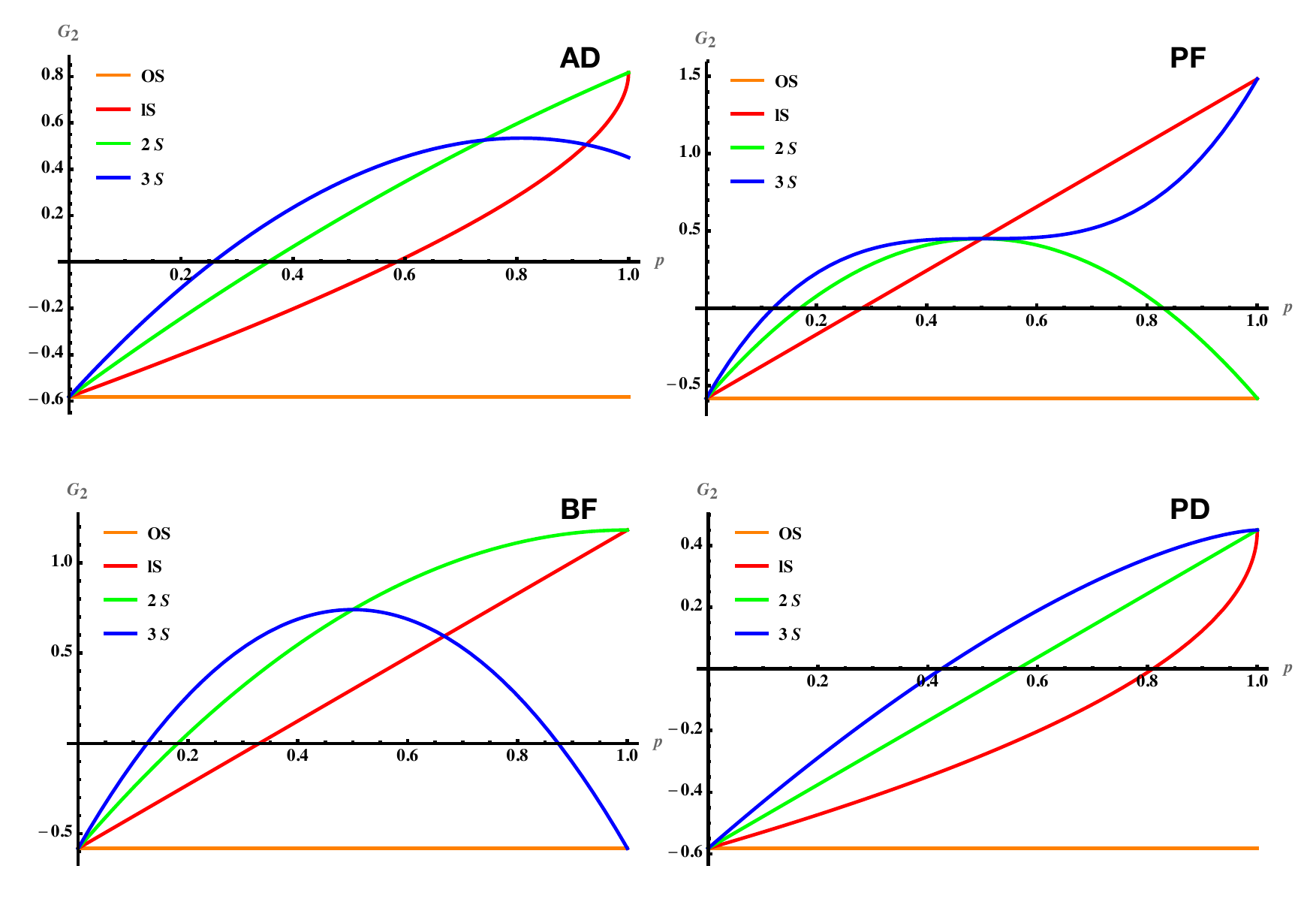}}
\caption{\footnotesize (Coloronline)  The variation of left hand side of the steering inequality (\ref{GHZ2}) (ordinate) for the post-processed GHZ state in (2 $\rightarrow$ 1) steering scenario versus the damping parameter 'p' (abscissa). Other notations are same as in fig. (\ref{fig7}).}  
\label{fig8}
\end{figure}

In this case only Charlie's subsystem is characterised. So Charlie performs the dichotomic POVMs given by (\ref{GHZsteer2}) on the initial state (\ref{GGHZ}). Charlie gets the target state as a GHZ state when the outcome is '0'. This can be seen by using Eq. (\ref{pCeqn}). The probability of success is $\text{Tr}[(\openone_2 \otimes \openone_2 \otimes P_0^C)\rho_{gGHZ}(\openone_2 \otimes \openone_2 \otimes P_0^{C^{\dagger}})] = 2 \sin^2 \theta$.
In this case also we manage to reveal hidden genuine steering of the gGHZ state ($\theta < 0.22$) by the action of the pre-processing operations. We now analyse the action of various damping channels on the post-processed state. 
\begin{enumerate}
	\item \textit{Amplitude damping:} The tendency of Alice to demonstrate genuine steering to Bob-Charlie decreases as the number of particles undergoing damping increases from one to three as shown in Fig. (\ref{fig8}).  The values of $\delta p_c^g$ are 0.16, 0.12, 0.09 that shows significant improvement in robustness after pre-processing.
	\item \textit{Phase flip:} The values of $\delta p_c^g$ are 0.03, 0.02, 0.02 and $\delta p_r^g$ is -0.02. This implies improvement in robustness, as well as, revival after pre-processing.
	\item \textit{Bit flip:} The values of $\delta p_c^g$ are 0.06, 0.03, 0.02 and $\delta p_r^g$ is -0.03 which is a significant improvement in robustness, as well as, revival after pre-processing.
	\item \textit{Phase damping:} In this case, the values of $\delta p_c^g$ are 0.06, 0.06, 0.05.
\end{enumerate}

Next, consider the case when shared state is a W-class state (\ref{wclass}) or one-parameter W-class state (\ref{owclass}).

\subsection{Pre-processing of W-class state}

In this case, the target is to obtain the W state. Unlike the case of gGHZ state where all the parties with characterised subsystem need not to do pre-processing in order to get the target state, here all parties with the characterised subsystem have to do pre-processing to get the target state. Let us first consider Alice to Bob-Charlie steering scenario.
\subsubsection{Alice to Bob-Charlie}
In this case Bob's and Charlie's subsystems are characterised. Bob and Charlie perform quantum measurements (POVM) with the following measurement operators on the initial state (\ref{wclass}).
\begin{eqnarray}
	P_{0}^{B} &=& \left(
\begin{array}{cc}
 \frac{c_1}{\sqrt{1-c_0^2-c_1^2}} & 0 \\
 0 & 1 \\
\end{array}
\right), \quad P_{1}^{B} =  \left(
\begin{array}{cc}
\sqrt{\frac{1-c_0^2-2c_1^2}{1-c_0^2-c_1^2}} & 0 \\
 0 & 0 \\
\end{array}
\right), \nonumber  \\
P_{0}^{C} &=& \left(
\begin{array}{cc}
 \frac{c_0}{\sqrt{1-c_0^2-c_1^2}} & 0 \\
 0 & 1 \\
\end{array}
\right), \quad P_{1}^{C} =  \left(
\begin{array}{cc}
 \sqrt{\frac{1-2c_0^2-c_1^2}{1-c_0^2-c_1^2}} & 0 \\
 0 & 0 \\
\end{array}
\right) .
\label{Wsteer1}
\end{eqnarray}
The target state is a W state when both the parties gets '0' outcome. This can be seen by using Eq. (\ref{pBCeqn}). The probability of success is $\text{Tr}[(\openone_2 \otimes P_0^B \otimes P_0^C)\rho_{GW}(\openone_2 \otimes P_0^{B^{\dagger}} \otimes P_0^{C^{\dagger}})] = \frac{3 c_0^2 c_1^2}{ 1-c_0^2-c_1^2}$. In this case also, we manage to reveal the hidden genuine steering of general W-class states by the action of pre-processing operations with some finite success probability. In order to calculate the improvement in the robustness after pre-processing, we will calculate $\delta p_c^w$ which is defined as the difference between the p values of the ($c_0 = 0.3, c_1 = 0.5$)/$(d_0 = 0.3)$ and ($c_0 = 1/\sqrt{3}, c_1 = 1/\sqrt{3})/(d_0 = 1/\sqrt{3})$ W-class state at which the genuine steerability of the state collapses. Also, improvement in the revival after pre-processing is observed by $\delta p_r^w$ which is defined as the difference between the p values of the ($c_0 = 0.3, c_1 = 0.5)/(d_0 = 0.3)$ and ($c_0 = 1/\sqrt{3}, c_1 = 1/\sqrt{3}$)/($d_0 = 1/\sqrt{3}$) W-class state at which the genuine steerability of the state revives. We now analyse the action of various damping channels on the post-processed state.

\begin{figure}[t!]
\resizebox{8cm}{7cm}{\includegraphics{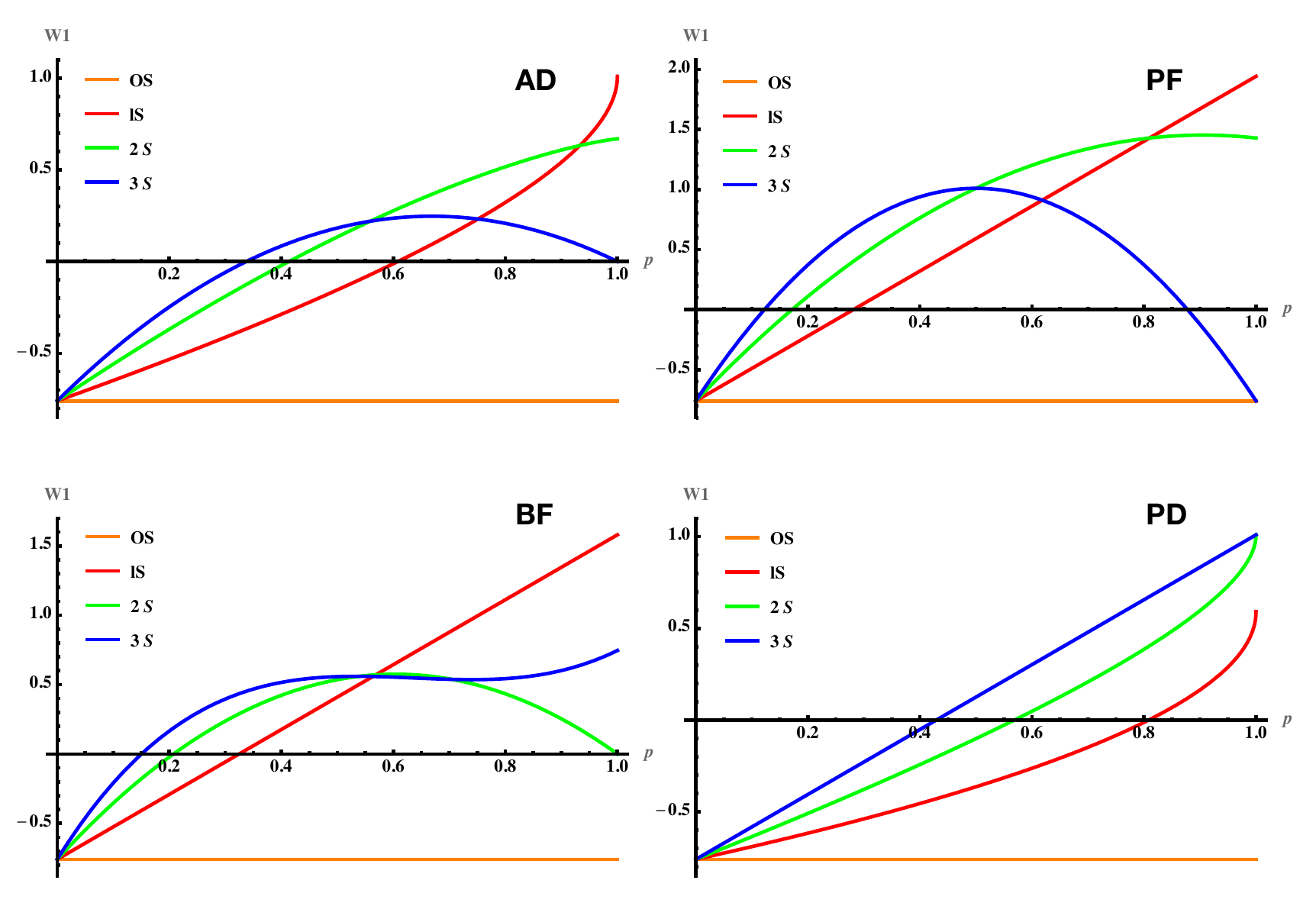}}
\caption{\footnotesize (Coloronline) The variation of left hand side of the genuine steering inequality (\ref{W1}) (ordinate) for the post-processed W-state in (1 $\rightarrow$ 2) steering scenario versus the damping parameter 'p' (abscissa). Other notations are same as in fig. (\ref{fig7}).} 
\label{fig9}
\end{figure}

\begin{enumerate}
	\item \textit{Amplitude damping:} We observed similar behaviour as before i.e. the tendency of Alice to demonstrate genuine steering to Bob-Charlie decreases as the number of particles undergoing damping increases from one to three as shown in Fig. (\ref{fig9}).  The values of $\delta p_c^w$ are 0.1, 0.09, 0.12 that shows significant improvement in robustness after pre-processing.
	\item \textit{Phase flip:} The values of $\delta p_c^w$ are 0.1, 0.06, 0.03 and $\delta p_r^w$ is -0.03. This implies improvement in robustness, as well as, revival after pre-processing.
	\item \textit{Bit flip:} The values of $\delta p_c^w$  are 0.14, 0.09, 0.05 which is a significant improvement in robustness.
	\item \textit{Phase damping:} In this case, the values of $\delta p_c^w$ are 0.21, 0.17, 0.11.
\end{enumerate}

Other features are same as before pre-processing. Next, consider the case when Alice-Bob demonstrate genuine steering to Charlie.

%\begin{figure}[!ht]
%\resizebox{6cm}{4cm}{\includegraphics{B14.pdf}}
%\caption{\footnotesize (Coloronline) $B^{14}_{max}$ is plotted against decoherence parameter $p$. The
%upper and lower curves represent Eq.(\ref{Bc1}) of case-\ref{c1} and Eq.(\ref{Bc2}) of case-\ref{c2}, respectively. The straight line shows the upper bound of bilocal correlations.}
%\label{b14}
%\end{figure}

\subsubsection{Alice-Bob to Charlie}
\begin{figure}[t!]
\resizebox{8cm}{7cm}{\includegraphics{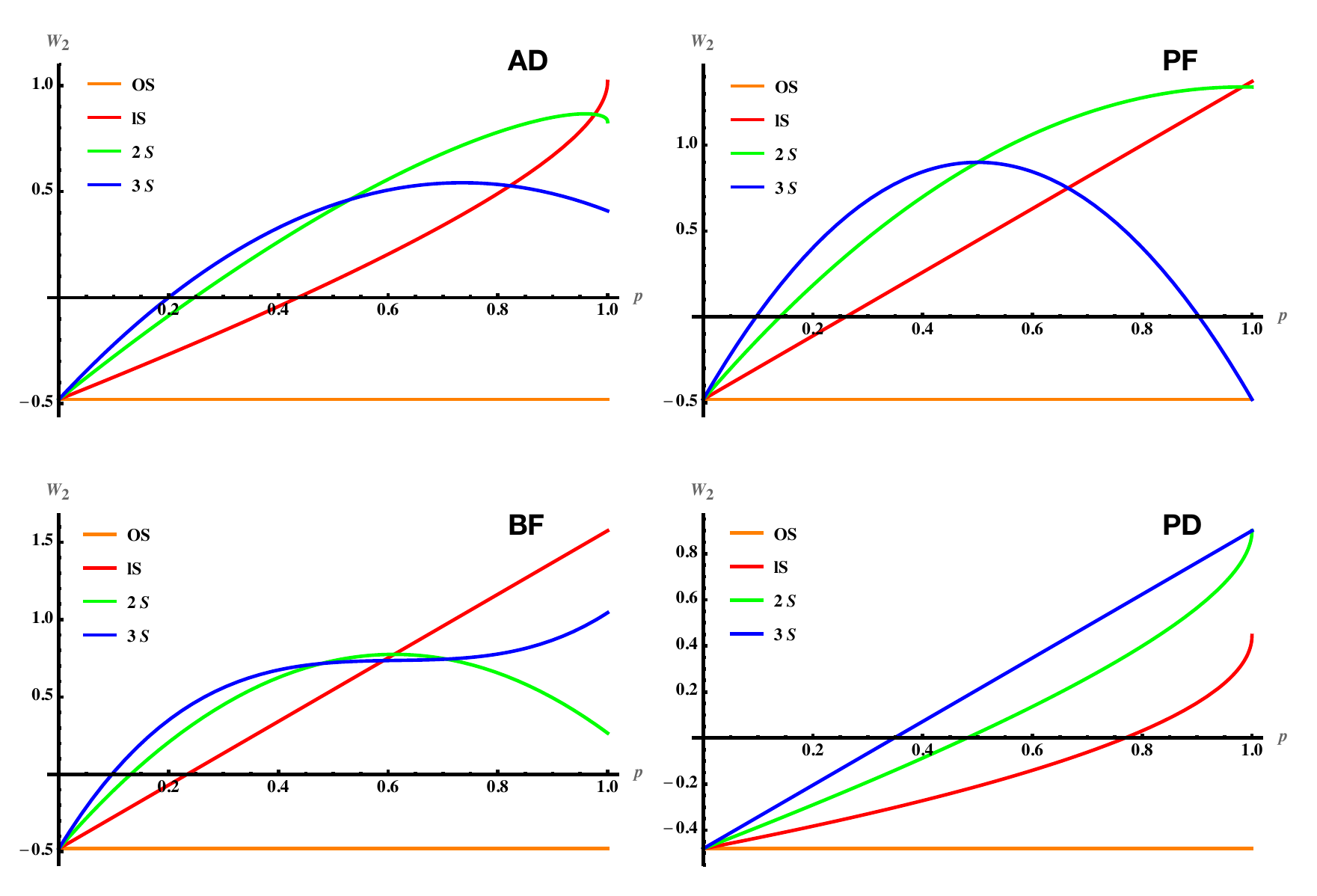}}
\caption{\footnotesize (Coloronline) The variation of left hand side of the genuine steering inequality (\ref{W2}) (ordinate) for the post-processed W-state in (2 $\rightarrow$ 1) steering scenario versus the damping parameter 'p' (abscissa). Other notations are same as in fig. (\ref{fig7}).}
\label{fig10}
\end{figure}

In this case only Charlie's subsystem is characterised. We have observed that there exists no dichotomic POVMs that Charlie can apply and gets W-state as a target state when the initial state is W-class state (\ref{wclass}). However, when the initial state is one-parameter W-class state (\ref{owclass}), Charlie perform quantum measurements (POVM) with the following measurement operators on the initial state (\ref{owclass}).
\begin{equation}
	P_0^C = \left(
\begin{array}{cc}
 \frac{\sqrt{2} d_0}{\sqrt{1-d_0^2}} & 0 \\
 0 & 1 \\
\end{array}
\right), \quad P_1^C = \left(
\begin{array}{cc}
 \sqrt{1- \frac{2 d_0^2}{1-d_0^2}} & 0 \\
 0 & 0 \\
\end{array}
\right)
\label{POVM2SW}
\end{equation}
The target state is a W state when Charlie gets '0' outcome. This can be seen by using Eq. (\ref{pCeqn}). The probability of success is $\text{Tr}[(\openone_2 \otimes \openone_2 \otimes P_0^C)\tilde{\rho}_{GW}(\openone_2 \otimes \openone_2 \otimes P_0^{C^{\dagger}})] = 3d_0^2$. In this case also, we manage to reveal the hidden genuine steering of one parameter W-class states ($d_0 < \frac{3}{25}$) by the action of pre-processing operations with some finite success probability. We now analyse the action of various damping channels on the post-processed state. 
\begin{enumerate}
	\item \textit{Amplitude damping:} The tendency of Alice to demonstrate genuine steering to Bob-Charlie decreases as the number of particles undergoing damping increases from one to three as shown in Fig. (\ref{fig10}).  The values of $\delta p_c^w$ are 0.12, 0.08 and 0.06 that shows significant improvement in robustness after pre-processing.
	\item \textit{Phase flip:} The values of $\delta p_c^w$ are 0.1, 0.06, 0.04 and $\delta p_r^w$ is -0.03. This implies significant improvement in robustness, as well as, revival after pre-processing.
	\item \textit{Bit flip:} The values of $\delta p_c^w$  are 0.09, 0.05 and 0.05 which is a significant improvement in robustness.
	\item \textit{Phase damping:} In this case, the values of $\delta p_c^w$ are 0.23, 0.17 and 0.11.
\end{enumerate}

\section{Summary and outlook} \label{5}

Understanding the effect of noise on the multipartite correlations are key in realising quantum informations tasks and future quantum technologies. In the present work we analyse the behaviour of genuine steering correlations \cite{Caval15,Caval16} involving three parties under the effect of noise modelled by the amplitude damping, phase flip, bit flip and phase damping. We consider two different genuine steering scenarios where either Alice demonstrates genuine steering to Bob-Charlie (2 $\rightarrow$ 1) or Alice-Bob together demonstrate genuine steering to Charlie (1 $\rightarrow$ 2). The three parties share either a genuinely steerable gGHZ state (\ref{GGHZ}) or a W-class state (\ref{wclass}), (\ref{owclass}). Genuine three qubit EPR steering is manifested through the violation of the respective steering inequality in a given scenario. We first show that the tendency to demonstrate genuine steering decreases as the number of parties undergoing decoherence increases from one to three. We have observed several instances where the genuine steerability of the state revives after collapsing if one keeps on increasing the damping and under asymmetric noise action of the dual channels. However, hidden genuine steerability of a state cannot be revealed solely from the action of noise. So, the parties having a characterised subsystem, perform local pre-processing operations depending upon the steering scenario and the state shared with the dual intent of revealing hidden genuine steerability or enhancing the same. In any potential applications of genuine steering correlations, e.g. multipartite secret sharing \cite{sss1} and quantum key distribution \cite{branciard} in quantum networks, many others\cite{Kimble2008,sss2,Mattar2017}, environmental noise naturally inhibits such correlations. The significance of our study lies in deciding the most robust state, determining the chances of revival and performing the exact pre-processing operations to reveal hidden genuine steering or enhancing the existing genuine steering while implementing such protocols.

Before concluding, it is worth mentioning certain possible offshoots of our work. First, practical demonstration of the exact pre-processing operations aimed towards revealing hidden genuine steering or enhancing it by photon based experiments should not be difficult to implement. Secondly, since weak measurement technique has been shown to shield the effect of amplitude damping for a number of quantum correlations, a systematic comparison between our pre-processing operations and weak measurements is worth doing in future. Finally, it would be interesting to investigate the effect of other procedures or devising new ones to further shield the effect of noise and finding optimum settings for the revival of correlations under the action of the noise.

{\it Acknowledgements:} SG acknowledges the S. N. Bose National Centre for Basic Sciences and QuNu Labs Pvt Ltd for the financial support. SG thanks Archan S Majumdar for the fruitful discussion. 

\textbf{Data availability statement-} The datasets generated during and/or analysed during the current study are available from the corresponding author on reasonable request.

\end{document}